\documentclass[pre,aps,10pt,superscriptaddress,twocolumn,floatfix]{revtex4-2}

\usepackage[utf8]{inputenc} 
\usepackage[nice]{nicefrac}
\usepackage{graphicx,color}
\usepackage{amsmath,amssymb,bm,bbm}
\usepackage{amsmath}
\usepackage{braket}
\usepackage{float}
\usepackage{placeins}
\usepackage{tabularx}

\usepackage{physics} 
\usepackage[normalem]{ulem} 
\usepackage[T1]{fontenc}
\usepackage[utf8]{inputenc}

\usepackage[plainpages=false,pdfpagelabels,colorlinks=true,linkcolor=red,urlcolor=blue,citecolor=blue,pdftitle={},pdfauthor={},pdfdisplaydoctitle=true,pdfduplex=DuplexFlipLongEdge]{hyperref}

\definecolor{darkred}{rgb}{0.90,0.2,0.2}
\definecolor{darkgreen}{rgb}{0,0.60,.2}
\definecolor{darkblue}{rgb}{0.1,0.3,1}
\definecolor{grey}{cmyk}{0,0,0,0.25}
\definecolor{orange}{cmyk}{0,0.6,0.8,0}

\usepackage[normalem]{ulem}

\begin{document}

\title{Tunable Hilbert-space fragmentation and extended critical regime}

\author{Mateusz Lisiecki}
\affiliation{Institute of Theoretical Physics, Wroclaw University of Science and Technology, 50-370 Wroc{\l}aw, Poland}
\author{Janez Bonča}
\affiliation{J. Stefan Institute, SI-1000 Ljubljana, Slovenia}
\affiliation{Faculty of Mathematics and Physics, University of Ljubljana, SI-1000 Ljubljana, Slovenia\looseness=-1}
\author{Marcin Mierzejewski}
\affiliation{Institute of Theoretical Physics, Wroclaw University of Science and Technology, 50-370 Wroc{\l}aw, Poland}
\author{Jacek Herbrych}
\affiliation{Institute of Theoretical Physics, Wroclaw University of Science and Technology, 50-370 Wroc{\l}aw, Poland}
\author{Patrycja Łydżba}
\affiliation{Institute of Theoretical Physics, Wroclaw University of Science and Technology, 50-370 Wroc{\l}aw, Poland}

\begin{abstract} 
Systems exhibiting the Hilbert-space fragmentation are nonergodic, and their Hamiltonians decompose into exponentially many blocks in the computational basis. In many cases, these blocks can be labeled by eigenvalues of statistically localized integrals of motion (SLIOMs), which play a similar role in fragmented systems as local integrals of motion in integrable systems.
While a nonzero perturbation eliminates all nontrivial conserved quantities from integrable models, we demonstrate for the $t$-$J_z$ chain that an appropriately chosen perturbation may gradually eliminate SLIOMs (one by one) by progressively merging the fragmented subspaces. This gradual recovery of ergodicity manifests as an extended critical regime characterized by multiple peaks of the fidelity susceptibility. Each peak signals a change in the number of SLIOMs and blocks, as well as an ultra-slow relaxation of local observables.
\end{abstract}

\maketitle

\section{Introduction.}
Closed quantum systems are generally ergodic. In such systems, initial states relax to equilibrium characterized by a few local conserved quantities or local integrals of motion (LIOMs), like energy or particle number~\cite{Rigol2008,Eisert_2015,Deutsch_2018}. Moreover, the fidelity susceptibility, i.e., the sensitivity of the Hamiltonian to perturbations, grows exponentially with the system size~\cite{Sierant_2019b,PhysRevX.10.041017,lim2024}. Nevertheless, there are nonergodic exceptions. The most studied examples are integrable systems that have an extensive number of LIOMs, with configurations of their eigenvalues labeling all energy eigenstates~\cite{Mukhin2009,Kirillov_2014}. As a result, these systems evolve to a steady state described by the generalized Gibbs ensemble~\cite{PhysRevLett.98.050405,PhysRevA.74.053616,Vidmar_2016}, and their fidelity susceptibility grows only polynomially with system size~\cite{PhysRevB.107.184312,Pozsgay_2024}.

The transition from integrable to nonintegrable dynamics in closed quantum systems is of broad interest, as it gives rise to a range of intriguing phenomena. For example, nearly integrable systems can exhibit prethermalization~\cite{Mori_2018,PhysRevB.84.054304,PhysRevLett.115.180601,PhysRevX.9.021027,PhysRevB.109.L161109,PhysRevLett.96.067202}, where their time evolution imitates integrable dynamics for finite times, being governed by local operators that are approximately conserved. Nearly integrable systems also display the so-called maximal chaos, signaled by the fidelity susceptibility larger than in ergodic systems and peaking at the integrability-breaking transition~\cite{PhysRevA.99.042117,PhysRevX.5.031007,PhysRevLett.103.170501,PhysRevB.77.245109,PhysRevA.77.032111,PhysRevLett.98.110601,PhysRevB.75.014439,LeBlond_2021,Sels_2021,świętek2025}. This peak indicates an ultra-slow relaxation of local operators.

Recently, a new phenomenon dubbed Hilbert-space fragmentation has been discovered~\cite{PhysRevX.10.011047,10.21468/SciPostPhys.15.3.093,PhysRevB.101.174204,PhysRevB.108.045127,Moudgalya_2021,PhysRevB.110.045418,PhysRevB.111.045411}. The Hamiltonian of such a system is block-diagonal in a simple local (experimentally accessible) basis. The number of its blocks grows exponentially with system size, and the dimension of the largest block remains a vanishing fraction of the total Hilbert space dimension, so that no dominant block emerges. Systems exhibiting fragmentation are nonergodic, but this behavior can be attributed to LIOMs in only few cases~\cite{PhysRevLett.132.220405,PhysRevE.104.044106}. Otherwise, it 
is believed to arise from statistically localized integrals of motion (SLIOMs)~\cite{PhysRevB.101.125126,PhysRevX.12.011050}. Interestingly, these operators are nonlocal, but they behave as local in typical states~\cite{PhysRevB.101.125126}. We emphasize that the differences between fragmented and integrable systems are not yet fully understood. The $t$-$J_z$ model~\cite{PhysRevB.55.6491,PhysRevB.36.381,Kotrla199033}, in which double occupancies of sites are prohibited, leading to kinetic constraints, serves as an exemplary model that exhibits fragmentation~\cite{PhysRevB.101.125126}.

In this manuscript, we introduce a protocol for tuning the conservation laws and the Hilbert-space fragmentation in the $t$-$J_z$ model. This simple protocol is based on additional hopping terms. It enables the control over the ergodic-nonergodic transition in such a way that it involves a sequence of transitions between distinct nonergodic regimes. We find that the introduced model, with its tunable fragmentation, supports an extended critical regime, in which the fidelity susceptibility exhibits multiple maxima. Each maximum indicates a change in the number of SLIOMs and blocks, and their number grows with system size. This stands in striking contrast to standard integrability-breaking transitions, which feature a single peak that signals erasing of (almost) all local conserved quantities~\cite{LeBlond_2021,Sels_2021,świętek2025}.

\begin{figure*}[htb!]
\includegraphics[width=\textwidth]{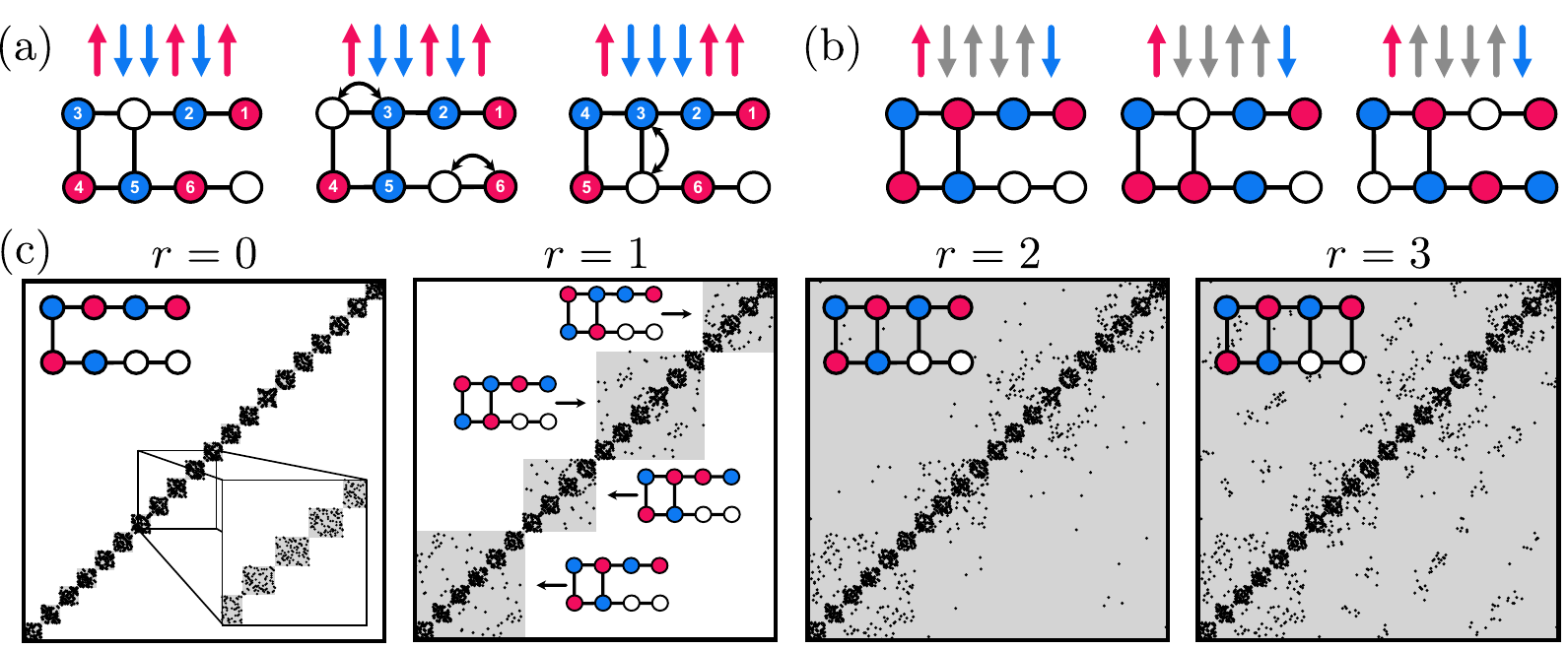}
\caption{A system with $L = 8$ sites, $N = 6$ particles and total spin $S^z = 0$.
(a) Illustration of two types of particle hops. \textcolor{black}{The spin pattern is indicated above the sketch of the ladder. Comparing the left and middle configurations, it is clear that the hops within the ladder legs do not modify the spin pattern. Specifically, the third and sixth particles, despite changing sites, remain the third and sixth particles. In contrast, comparing the left and right configurations reveals that the hops between the ladder legs do modify the spin pattern. The former fifth particle becomes the third, the former third becomes the fourth, and the former fourth becomes the fifth.
(b) Illustration of SLIOMs. The particles at the edges of the ladder always have the same index, and so correspond to the same spin in the spin pattern, as they cannot hop to the opposite leg. They are shown with blue and red arrows. Other spins in the spin pattern are not fixed and can be flipped, as illustrated by the exemplary configurations in the middle and right sketches. They are marked with grey arrows.}
(c) The variation of the Hamiltonian matrix in the computational basis with the number of rungs $r$. State indices increase from left to right and from bottom to top. Grey squares mark blocks. For $r = 1$, an exemplary state from each block is shown. For $r \ge 2$, there is a single block, but clusters of matrix elements become more interconnected as $r$ increases. }
\label{sketch}
\end{figure*}

\section{$t$-$J_z$ model.}
In the following, we study the infinite-interaction limit of the anisotropic Fermi-Hubbard Hamiltonian, i.e., the so-called $t$-$J_z$ model~\cite{PhysRevB.55.6491,PhysRevB.36.381,Kotrla199033}. We consider the Hilbert space that incorporates the hard-core constraint, where the on-site basis consists of three states: $0, \uparrow, \downarrow$. In this Hilbert space, the Hamiltonian of the $t$-$J_z$ model can be written as
\begin{equation}
\label{eq:H_tJz}
    H=-t\sum_{i=1}^{L-1}\sum_{\sigma=\uparrow,\downarrow}\left({c}_{i,\sigma}^\dagger {c}_{i+1,\sigma}+\text{H.c.}\right)+J_{z}\sum_{i=1}^{L-1}S_{i}^{z}S_{i+1}^z,
\end{equation}
where $L$ denotes the number of lattice sites, ${c}^\dagger_{i,\sigma}$ (${c}_{i,\sigma}$) creates (annihilates) a fermion with spin $\sigma$ at site $i$, while $S_i^z = 1/2({c}_{i,\uparrow}^\dagger, {c}_{i,\downarrow}^\dagger) \sigma^z ({c}_{i,\uparrow}, {c}_{i,\downarrow})^T$ is the spin operator with $\sigma_z$ corresponding to the Pauli matrix. This Hamiltonian conserves the total particle number, $N = N_\uparrow + N_\downarrow$, as well as the total spin projection, $S^z = \frac{1}{2} (N_\uparrow - N_\downarrow)$, where $N_\sigma = \sum_{i=1}^{L} {c}_{i,\sigma}^\dagger {c}_{i,\sigma}$. For simplicity, we restrict considerations to a single symmetry sector with a defined $N$ and $S^z$. Its dimension is given by $Z=\binom{L}{N}\binom{N}{N_\sigma}$.

A hallmark of the $t$-$J_z$ model is its kinetically constrained dynamics, which consists of the reordering of $N_h = L - N$ holes so that the overall spin pattern remains frozen. \textcolor{black}{(The spin pattern corresponds to the sequence of spins obtained after removing all holes. While the spin of the $k$th site is not conserved, the spin of the $k$th particle is.)} As a result, the Hamiltonian exhibits the Hilbert-space fragmentation~\cite{PhysRevB.101.125126} and decomposes into disconnected blocks in the computational basis. There are $B = \binom{N}{N_\sigma}$ such blocks, each of dimension $\tilde{Z} = \binom{L}{N}$.

We recall that in integrable systems, configurations of eigenvalues of LIOMs uniquely label all energy eigenstates~\cite{faddeev1996}. In a similar spirit, it has been proposed that in fragmented systems, configurations of eigenvalues of SLIOMs uniquely determine all blocks~\cite{PhysRevB.101.125126,PhysRevX.12.011050}. For the $t$-$J_z$ model, these SLIOMs have already been established, and each of them corresponds to the spin of the $k$th particle with $k\le N$. It is worth emphasizing that they are not local in the operator sense, i.e., they cannot be written as sums of operators with a finite support over the lattice. Nevertheless, they are local in the statistical sense: the $k$th particle can move through the lattice, but its motion is restricted by the number of holes. Therefore for typical computational basis states, this particle can be found in the vicinity of the $k L / N$ site~\cite{PhysRevB.101.125126}. We highlight that the knowledge of SLIOMs is sufficient to characterize the dynamics of a fragmented system, for example, by providing a tight Mazur bound~\cite{PhysRevB.101.125126,PhysRevX.12.011050}.

\section{Tunable Hilbert-space fragmentation.}
We now demonstrate that it is possible to tune the fragmentation by progressively reducing the number of SLIOMs and, consequently, the number of blocks. This can be achieved in two steps and requires $N_h\ll N$. First, we deform a chain with $L$ sites, forming a two-leg ladder with $L/2$ sites coupled by a single rung between positions $i = L/2$ and $j = L/2 + 1$ (see Fig.~\ref{sketch}). \textcolor{black}{Consequently, the coordination number of sites $i$ and $j$ becomes larger ($N_Z>2$) than that of the other sites ($N_Z=2$).} Site and particle indices are assigned starting from the rightmost position in the upper leg of the ladder. Next, we progressively add additional $r$ rungs between positions $i = L/2 - l$ and $j = L/2 + 1 + l$ with \mbox{$l=1,\dots, r$}. Explicitly, each additional rung corresponds to $h(l)=-t(l)\sum_{\sigma=\uparrow,\downarrow}({c}_{i,\sigma}^\dagger {c}_{j,\sigma}+\text{H.c.})+J_{z}(l)S_{i}^{z}S_{j}^z$, where both $t(l)$ and $J_{z}(l)$ can depend on the rung index, $l$. \textcolor{black}{Let us consider the following scenario: a particle reaches a site with $N_Z>2$ in the lower leg, see the left panel of Fig.~\ref{sketch}(a).} If there is a hole in the upper leg, the particle can hop and modify the spin sequence, as illustrated in the right panel of Fig.~\ref{sketch}(a). \textcolor{black}{Therefore such a particle can hop both along the legs and across the rungs. Since it is able to move through sites with $N_Z>2$, its motion is effectively two-dimensional. In contrast, there are particles located far from the rungs, i.e., those that either cannot reach any rung or can reach only the last rung when no holes are present in the opposite leg. Their movement is then restricted to sites with $N_Z=2$, so it remains effectively one-dimensional. Notably, particles moving only through sites with $N_Z=2$ correspond to the same spin in the spin pattern, whereas those that are able to move through sites with $N_Z>2$ do not. Consequently, the dynamics of the former cannot affect the spin pattern, and they continue to generate SLIOMs. The described mechanism leads to the merging of blocks. However, not all blocks are merged, since those that differ in the spins of particles moving only through sites with $N_Z=2$ remain decoupled.}

The number of SLIOMs in the above setup can be estimated as follows. Consider separating holes by moving them to the rightmost positions in the lower leg, see the left panel of Fig.~\ref{sketch}(b). This reveals that there are exactly $N-L/2-r=L/2-N_h-r$ particles in the lower leg that have fixed spins. This reasoning can be reversed, yielding the same number of particles in the upper leg. Therefore the total number of SLIOMs is given by $Q=L-2N_h-2r$. If $Q\le N_\sigma$ for both $\sigma$, SLIOMs may have arbitrary spin projections, and the number of blocks can be calculated as $B=\sum_{n=0}^{Q} \binom{Q}{n}=2^Q$.
Otherwise, if $Q >  N_\sigma$ for at least one $\sigma$, certain spin configurations are inaccessible and one should adjust the range of summation. Then, the general expression for the number of blocks reads
\begin{equation}
B=\sum_{n=\text{max}[0,Q-N_{-\tilde{\sigma}}]}^{\text{min}[Q,N_{\tilde{\sigma}}]} \binom{Q}{n},
\label{bnumber}
\end{equation}
where $\tilde{\sigma}$ is the spin projection of the minority species. Simultaneously, the dimension of a single block depends on the number of SLIOMs that have a spin projection $\tilde{\sigma}$, which is denoted by $n$. This dimension can be determined by counting all possible ways to distribute $N$ particles across $L$ sites and \textcolor{black}{to distribute $N_{\tilde{\sigma}}-n$ minority spins across $N - Q$ particles that can move through sites with $N_Z>2$, i.e., $\tilde{Z}=\binom{L}{N}\binom{N-Q}{{N}_{\tilde{\sigma}}-n}$.} \textcolor{black}{We note that the analytical expression for the number of blocks from Eq.~\eqref{bnumber} is valid for arbitrary parameters $L$, $N$ (or $N_h$) and $N_{\tilde{\sigma}}$ (or $N_{-\tilde{\sigma}}$). Therefore it remains valid also in the thermodynamic limit.}

In integrable systems, a perturbation or a change in lattice dimensionality destroys (almost) all LIOMs. In contrast, the system considered above exhibits, to the best of our knowledge, the only known example of tunable Hilbert-space fragmentation. Specifically, it allows controlling the number of conservation laws, as adding a single rung to a ladder with $r > 0$ breaks exactly two SLIOMs. Our numerical calculations did not reveal any additional blocks $B$ beyond those introduced in Eq.~(\ref{bnumber}). In Fig.~\ref{sketch}(c), we illustrate the described behavior: adding one rung to a ladder with $L = 8$, $N=6$ and $r=0$ reduces the number of blocks from $20$ to $4$, and adding another rung reduces it further from $4$ to $1$. Additional examples illustrating the tunability of fragmentation are provided in \textcolor{black}{App.~\ref{app:a}}.

\textcolor{black}{We highlight that the main reason for selecting the \mbox{$t$-$J_z$} model, beyond fragmentation, is that blocks are ergodic (with the exception of those with a completely ferromagnetic and antiferromagnetic spin pattern~\cite{PhysRevB.101.125126}). Consequently, the gradual merging of these blocks corresponds to the progressive recovery of ergodicity for any $J_z\neq 0$. The specific value of $J_z$ does not play a significant role, as it does not affect the number of SLIOMs, $Q$, and consequently the number of blocks, $B$. The situation changes when $J_z=0$, as the model becomes integrable, making it difficult to disentangle effects arising from fragmentation (SLIOMs) and integrability (LIOMs).}

We now confirm that the number of SLIOMs is reflected in the spin dynamics with direct numerical calculations. We consider a ladder with $t = J_z=1$ and $L = 16$, fixing $N=13$ and $S^z=1/2$. For the additional $r$ rungs, we also set $t(l) = J_z(l) = 1$. As the initial state, we choose the computational basis state $|\psi_0\rangle = \left|\uparrow\dots\uparrow 0 \dots 0\downarrow \dots \downarrow \rangle\right.$, where all spin-up (spin-down) particles occupy the rightmost positions in the upper (lower) leg. The state at finite times, $|\psi_t\rangle$, is obtained using the Lanczos propagation method~\cite{Lanczos_1950,Park_1986}. Finally, we compute the expectation values of spin projections, $\langle \psi_t | S^z_i | \psi_t \rangle$, for all sites $i \le L$.

\begin{figure}[htb!]
\includegraphics[width=\columnwidth]{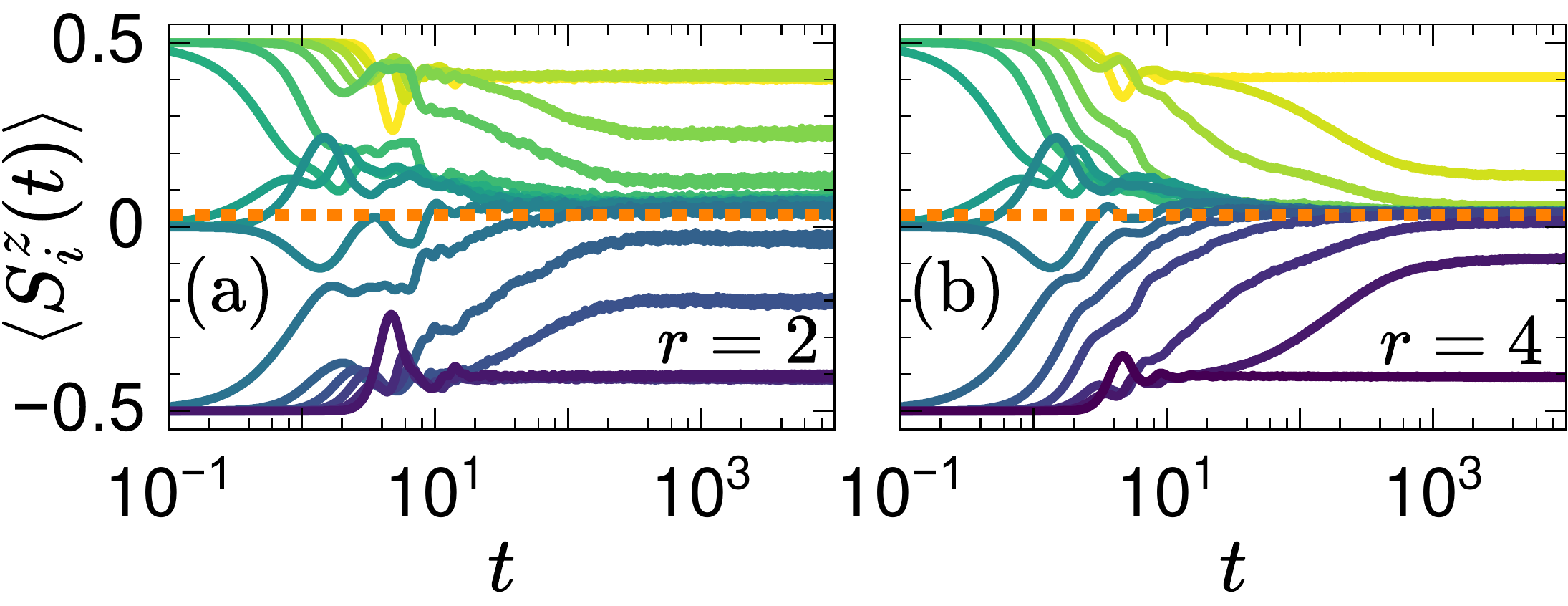}
\caption{The time evolution of spin projections, $\langle \psi_t | S^z_i | \psi_t \rangle$, for all sites $i\le L$. We focus on a system with $L=16$ sites, $N=13$ particles, and total spin $S^z=1/2$. Results in (a) and (b) were calculated for $r=2$ ($Q=6$) and $r=4$ ($Q=2$), respectively. The initial state was the computational basis state with spin-up (spin-down) particles occupying the rightmost positions in the upper (lower) leg. The dashed lines mark the thermal prediction $(N_\uparrow-N_\downarrow)/(2L)\approx 0.031$.}
\label{figjanez}
\end{figure}

Results for a ladder with $r=2$ ($Q=6$) and $r = 4$ ($Q=2$) are presented in panels (a) and (b) of Fig.~\ref{figjanez}, respectively. Changing the number of SLIOMs clearly modifies the expectation values $\langle \psi_t | S^z_i | \psi_t \rangle$. This conclusion is nontrivial, as nonlocal conservation laws typically do not affect the time evolution of local operators~\cite{PhysRevLett.124.040603}. It agrees with a recent observation that SLIOMs can give rise to a tight Mazur bound~\cite{PhysRevB.101.125126,PhysRevX.12.011050}, and it facilitates the study in the next section, where we treat SLIOMs in a similar spirit to LIOMs. We also highlight an exceptionally broad range of relaxation times of $\langle \psi_t | S^z_i | \psi_t \rangle$, a feature that is difficult to realize in integrable or nearly integrable models~\cite{Mierzejewski_2022,PhysRevB.92.195121,PhysRevResearch.5.043019}. In general, $S^z_i$ closer to the ladder edges exhibit larger projections onto SLIOMs and evolve more slowly. Some of these spins do not relax to the thermal prediction, although they are not strictly conserved.

\textcolor{black}{Before proceeding, we address the possibility of progressively including different hops between sites that do not correspond to the rungs considered above. Two important comments should be made. First, we cannot provide analytical expressions for hops between arbitrarily chosen pairs of sites. Second, one can always ask the same question as we did before: are there particles that can move only through sites with $N_Z=2$? If this is the case, their spins correspond to SLIOMs and determine the lower bound for the number of blocks. The strength of our original protocol is that this lower bound coincides with the exact number of blocks, as particles moving only through sites with $N_Z=2$ give rise to the only relevant conservation laws. This is not guaranteed when different hops between sites are considered.}

\section{Extended critical regime.}
Many measures are sensitive to the breaking of integrability and can therefore be used to detect the transition to the ergodic regime. By far, the most commonly studied ones are the spectral statistics~\cite{Santos_2010,Fogarty_2021,Szasz_2021,Sierant_2019,PhysRevB.88.115126}, entanglement entropy~\cite{PhysRevE.89.012125,Beugeling_2015,PhysRevE.100.062134,PhysRevLett.119.110604,PhysRevResearch.6.023030} and (recently) fidelity susceptibility~\cite{PhysRevA.99.042117,PhysRevX.5.031007,PhysRevLett.103.170501,PhysRevB.77.245109,PhysRevA.77.032111,PhysRevLett.98.110601,PhysRevB.75.014439}. Most of these measures are essentially binary in nature. For example, the spectral statistics can either follow Wigner-Dyson or Poisson distribution, while the entanglement entropy can either be maximal or not. Surprisingly, we now demonstrate that, in our ladder with tunable fragmentation, the fidelity susceptibility can reveal the breaking of only a few conservation laws, thereby identifying the transition between two nonergodic regimes.

The fidelity susceptibility of a single energy eigenstate, $H(\lambda)|n(\lambda)\rangle = E_n(\lambda)|n(\lambda)\rangle$, quantifies the sensitivity of this state to changes in the Hamiltonian, $\partial_\lambda H$. Its average over all energy eigenstates, $|n(\lambda)\rangle$, is given by~\cite{You_2007,Kolodrubetz_2017,PhysRevX.10.041017}
\begin{equation} 
\label{eq:Chi_av} 
\chi_\text{av} = \frac{1}{Z} \sum_{n=1}^{Z} \sum_{m \neq n} \frac{\omega_{nm}^2}{(\omega_{nm}^2 + \mu^2)^2} \left| \langle n (\lambda) | \partial_\lambda H | m  (\lambda)\rangle \right|^2, 
\end{equation}
where we defined $\omega_{nm} = E_n(\lambda) - E_m(\lambda)$. From now on, we focus on the limit $\lambda\rightarrow 0$ and make the dependence on $\lambda$ implicit. In Eq.~\eqref{eq:Chi_av}, we introduced the energy cutoff $\mu$, which is known as regularization~\cite{PhysRevX.10.041017}. Usually, $\omega_{H}<\mu\ll\Delta E$, where $\omega_H\propto 1/Z$ is the Heisenberg energy (the mean level spacing, serving as the minimal relevant energy scale in a finite system) and $\Delta E\propto L$ is the energy bandwidth. The role of $\mu$ is twofold. Firstly, it removes the issue of an ill-defined denominator in Eq.~\eqref{eq:Chi_av} for degenerate energy eigenstates. However, its true importance is revealed when the average fidelity susceptibility, $\chi_\text{av}$, is rewritten as~\cite{Kim_2024}
\begin{equation}
\label{eq:Chi_final}
    \chi_\text{av}\mu \propto |f(\omega \approx \mu)|^2,
\end{equation}
where the spectral function $|f(\omega)|^2$ is the Fourier transform of the \textcolor{black}{connected} autocorrelation function of $\partial_\lambda H$,
\begin{equation}
\label{eq:spectral}
    |f(\omega)|^2=\frac{1}{2\pi Z}\sum_{n=1}^{Z}\int_{-\infty}^{\infty} dt e^{i\omega t}\langle n|\partial_\lambda H(t)\partial_\lambda H|n\rangle_c,
\end{equation} 
and $\langle n|O(t)O|n\rangle_c=\langle n|O(t)O|n\rangle-\langle n|O(t)|n\rangle\langle n|O|n\rangle$.
The derivation of the relation from Eq.~\eqref{eq:Chi_final} can be found in \textcolor{black}{Appendix~\ref{app:b}}.
Thus studying the rescaled fidelity susceptibility $\chi_\text{av} \mu$ for different energy cutoffs $\mu$ amounts to probing the system dynamics at different times, $t \approx \mu^{-1}$. When the autocorrelation function of $\partial_\lambda H$ decays exponentially with the relaxation time $\tau$, the spectral function from Eq.~\eqref{eq:spectral} is a Lorentzian, whose maximum value is proportional to $\tau$. Therefore, for small energy cutoffs, $\mu \approx \omega_H$, the rescaled fidelity susceptibility, $\chi_\text{av} \mu$, provides a rough estimate of the relaxation time, $\tau$.

In ergodic systems, the spectral function $|f(\omega)|^2$ is related to the envelope function from the ETH ansatz and forms a plateau for low $\omega$~\cite{D_Alessio_2016}. This plateau indicates when the ETH ansatz becomes equivalent to the RMT ansatz in the frequency domain or, equivalently, when the perturbation $\partial_\lambda H$ relaxes in the time domain. Consequently, $\chi_\text{av} \mu$ is independent of $\mu^{-1}$, as obvious from Eq.~(\ref{eq:Chi_final}). In integrable systems, the spectral weight accumulates at $\omega = 0$, forming a Drude-like peak, and the spectral function $|f(\omega)|^2$ develops a gap at low $\omega\neq0$~\cite{Brenes_2020b,Brenes_2020,PhysRevE.102.062113}. This behavior can be related to the existence of LIOMS~\cite{PhysRevLett.124.040603,Vidmar_2021} and results in $\chi_\text{av} \mu$ decreasing with $\mu^{-1}$~\cite{lim2024,Kim_2024}. Near the ergodicity-breaking transition, the spectral function $|f(\omega)|^2$ usually develops a polynomially decaying tail that can span several orders of magnitude in $\omega$~\cite{lim2024,Kim_2024}. Such behavior indicates that LIOMs acquire finite relaxation times, leading to the broadening of a Drude-like peak~\cite{Vidmar_2021}. As a result, $\chi_\text{av} \mu \propto \mu^{-\alpha}$ with $\alpha \in (0,1]$. It exhibits a maximum pinned to the critical point, indicating the exceptionally slow relaxation of the perturbation $\partial_\lambda H$~\cite{LeBlond_2021,Sels_2021,świętek2025}. This phenomenon is known as maximal chaos.

We design the following experiment using the ladders described above with $t=J_z=1$. We treat $r$ as a continuous parameter, where its integer part \textcolor{black}{$[r]$} determines the number of additional rungs with the same coupling strength, i.e., $t(l)=J_z(l) = 1$ for $l = 1, \dots, [r]$. Simultaneously, its fractional part \textcolor{black}{$\{r\}$} sets the coupling strength of the last rung, i.e., $t(l) =J_z(l) = \left\{r\right\}$ for $l = [r] + 1$. This allows for the possibility of weakly broken SLIOMs for small but nonzero $\left\{r\right\}$, resembling weakly broken LIOMs in nearly integrable systems. We focus on the perturbation in the form of the next-nearest neighbor hopping, $\partial_\lambda H=\frac{1}{\sqrt{L}} \sum_{\sigma = \uparrow, \downarrow} \sum_{i=1}^{L-2} ( c_{i, \sigma}^\dagger c_{i+2, \sigma} + \text{H.c.})$, which is local, involves all lattice sites, and does not share the block structure of the Hamiltonian. The coefficient in front of the sum ensures the unit Hilbert-Schmidt norm, see the discussion in~\cite{lydzba2024}. Moreover, we focus on the middle of the spectrum, so that we average the rescaled fidelity susceptibility over a fraction of all energy eigenstates. \textcolor{black}{Technical details are discussed in Appendix~\ref{app:c}, while numerical results for additional perturbations are provided in App.~\ref{app:d}}. Finally, we tune $r$ and determine if $\chi_\text{av}\mu$ develops maxima for vanishing $\left\{r\right\}$.

\begin{figure}[t!]
\includegraphics[width=\columnwidth]{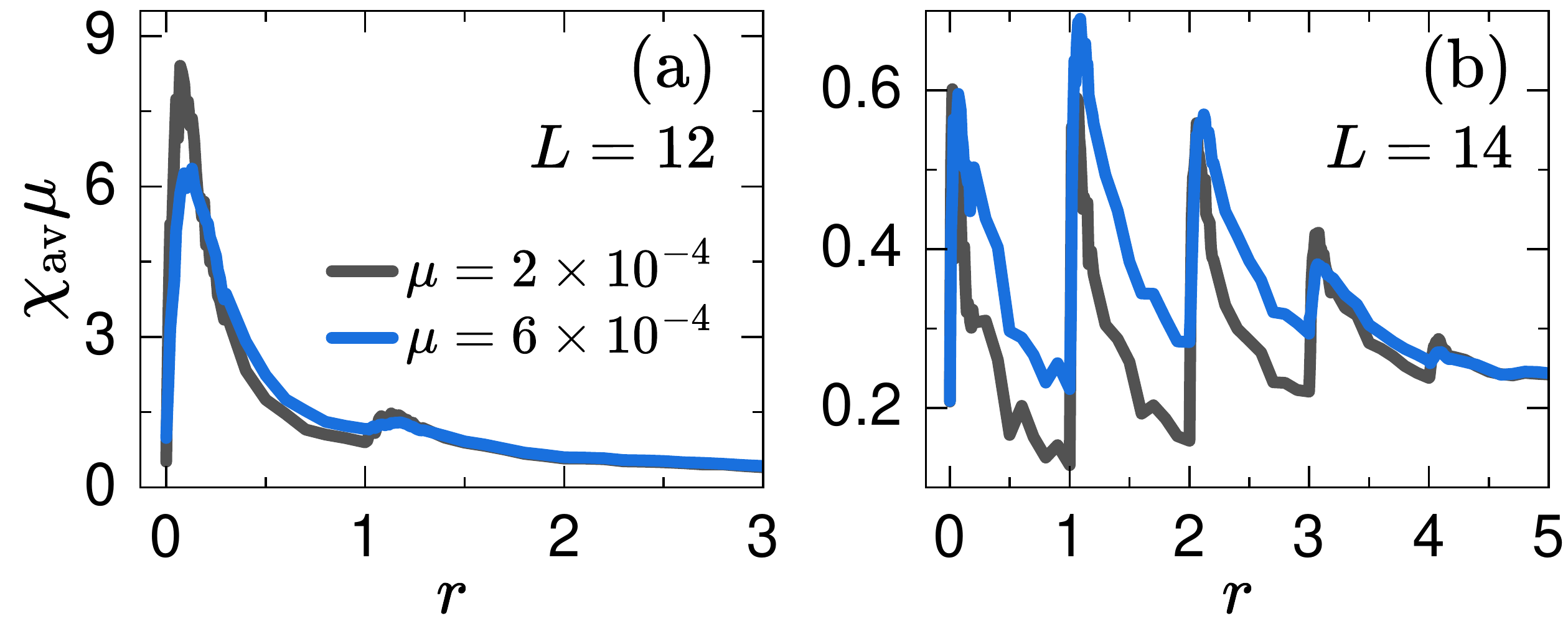}
\caption{Rescaled fidelity susceptibilities $\chi_\text{av}\mu$ plotted against rung number $r$ for a ladder with (a)~$L=12$ sites, $N=8$ particles and total spin $S^z=0$. Analogical results, but for a ladder with $L=14$ sites, $N=12$ particles, and total spin $S^z=0$, are presented in (b). We consider two energy cutoffs, $\mu = 2 \times 10^{-4}$ and $6 \times 10^{-4}$.}
\label{fig1}
\end{figure}

\begin{figure}[t!]
\includegraphics[width=\columnwidth]{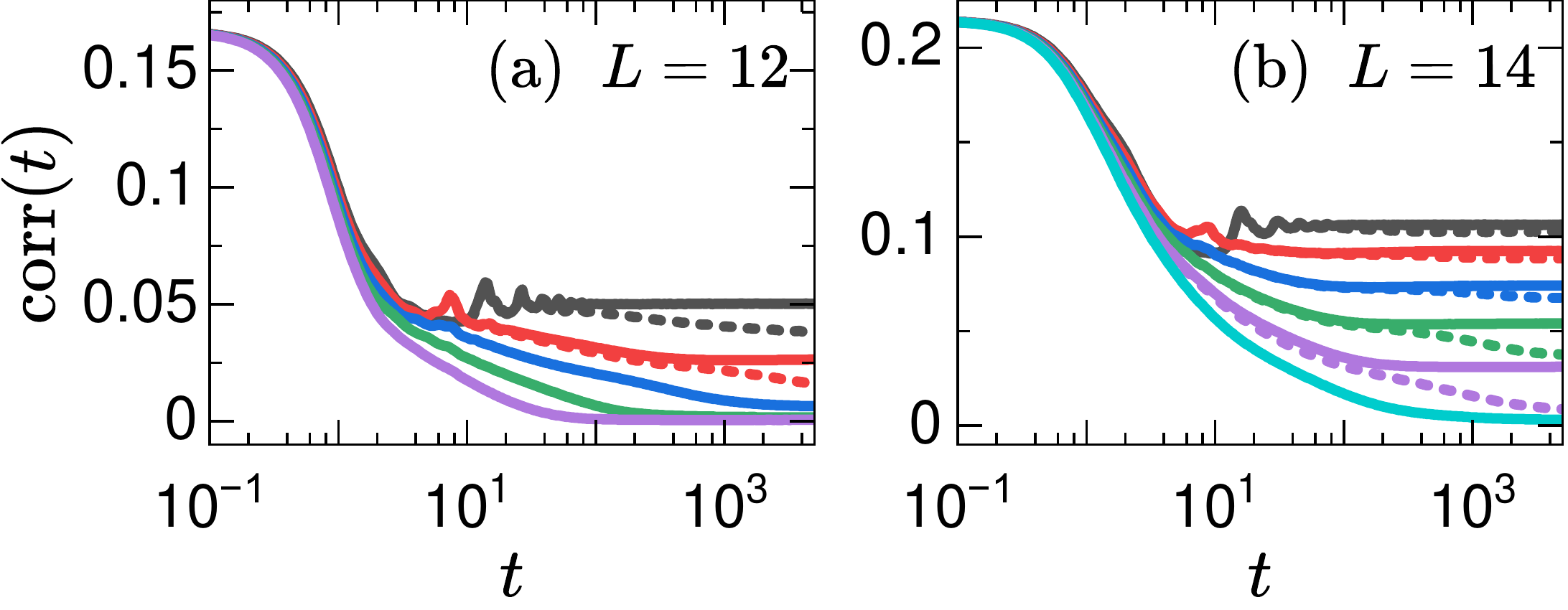}
\caption{\textcolor{black}{(a) The spin-spin autocorrelation function $\text{corr}(t)$ for a ladder with $L=12$ sites, $N=8$ particles, and total spin $S^z=0$. Analogous results for a ladder with $L=14$ sites, $N=12$ particles, and total spin $S^z=0$ are shown in (b). The longest considered time corresponds to the Heisenberg time, $\tau_H=\omega_H^{-1}$, obtained from the entire energy spectrum. Different solid lines correspond to different numbers of additional rungs, $r \in \{0,1,2,3,4\}$ for $L=12$ and 
$r \in \{0,1,2,3,4,5\}$ for $L=14$.
The long-time value of $\text{corr}(t)$ decreases with $r$. Recall that $\chi_\text{av}^\mu$ does not exhibit a peak at integer $r$. Dashed lines were calculated for fractional $r$ near the peak of $\chi_\text{av}^\mu$, i.e., $r \in \{0.1,1.2\}$ in (a) and $r \in \{0.05,1.05,2.05,3.1,4.2\}$ in (b). Here, $\text{corr}(t)$ exhibits an ultra-long relaxation time, $\tau \propto \tau_H$.}}
\label{fig_corr_t}
\end{figure}

In panels (a) and (b) of Fig.~\ref{fig1}, we show the evolution of the rescaled fidelity susceptibility, $\chi_\text{av}\mu$, with the number of rungs, $r$, for a ladder with $L = 12\,, N=8,S^z=0$ and $L = 14\,, N=12,S^z=0$, respectively. For both ladders and energy cutoffs considered, $\chi_\text{av}\mu$ develops pronounced peaks, which signal an ultra-slow relaxation of $\partial_\lambda H$, near all transitions that separate regimes with different numbers of SLIOMs. This behavior is nontrivial, as it demonstrates that $\chi_\text{av}\mu$, unlike other standard ergodicity indicators, is sensitive to the breaking of as few as two SLIOMs. \textcolor{black}{In App.~\ref{app:e}}, we also confirm that $\chi_\text{av}\mu$ increases with $\mu^{-1}$ near transitions (for $\left\{r\right\} \approx 0$), while it decreases with $\mu^{-1}$ away from transitions (for $\left\{r\right\} \gg 0$). Consequently, the critical regime extends over $0 \leq r \leq r'$ and exhibits $r'+1$ peaks, where fragmentation disappears at $r'=L/2-N_h$. The number of such peaks increases with the system size, since more rungs can be added before all SLIOMs are broken. This is remarkable, as conventional integrability-breaking transitions, like the one induced by adding the next-nearest-neighbor hopping to the Heisenberg XXZ chain~\cite{LeBlond_2021}, feature a single peak. The existence of an extended critical regime with multiple peaks of $\chi_\text{av}\mu$ is one of the most important outcomes of our study.

\textcolor{black}{Inspired by the results on the rescaled fidelity susceptibility, we investigate a related quantity, namely the autocorrelation function. Its connected version is defined in Eq.~\eqref{eq:spectral}. Here, we focus on the spin-spin autocorrelation function summed over all lattice sites:
\begin{equation}
    \text{corr}(t) = \frac{1}{ZL} \sum_{n=1}^{Z} \sum_{i=1}^{L} \langle n | S_i^z(t) S_i^z | n \rangle .
\end{equation}
However, the same qualitative behavior is expected for autocorrelation functions of other local operators.}

\textcolor{black}{In Fig.~\ref{fig_corr_t}(a), we plot $\text{corr}(t)$ for for a ladder with $L=12$ sites, $N=8$ particles, and total spin $S^z=0$. Analogous results for a ladder with $L=14$ sites, $N=12$ particles, and total spin $S^z=0$ are shown in Fig.~\ref{fig_corr_t}(b). When $Q>0$ and $B>1$ ($r\le 1$ for $L=12$ and $r\le 4$ for $L=14$), the long-time value of the spin-spin autocorrelation function is far from thermal, but it decreases as more rungs are added (see solid lines, where $r$ is integer and $\chi_\text{av}\mu$ shows no peak). Moreover, the change in the number of rungs is first signaled by an ultra-long relaxation of $\text{corr}(t)$ (see dashed lines, where $r$ is fractional and $\chi_\text{av}\mu$ exhibits a peak). These observations suggest that SLIOMs may influence the dynamics similarly to LIOMs. Specifically, breaking a SLIOM with a nonzero projection onto a given $S_i^z$ decreases diagonal matrix elements of this $S_i^z$, thereby reducing the long-time value of $\text{corr}(t)$. 
However, as demonstrated in the preceding sections, SLIOMs, unlike LIOMs, can be eliminated one by one. Consequently, successive breaking of SLIOMs {\em gradually} restores the long-time value of $\text{corr}(t)$ toward its thermal prediction, i.e., $\text{corr}(t \to \tau_H)\to 0$ (up to finite-size corrections of order $N^2/L$). In the context of integrability, the thermal value is expected already for arbitrarily small perturbations, although the corresponding relaxation time can be ultra-long.} 

\textcolor{black}{\section{Thermodynamic limit.}}
\textcolor{black}{Before concluding, we comment on how to approach the thermodynamic limit and on the behavior of the rescaled fidelity susceptibility in this limit. In the conventional scenario, in which the number of lattice sites $L$ is increased while keeping the particle density $\rho$ fixed, we obtain the following expression:
\begin{equation}
Q = (2\rho - 1)L - 2r,
\end{equation}
where $r \neq 0$ and $N/L = \rho = \text{const.}$ It seems the most natural to scale the number of rungs $r$ in the same way as the particle number $N$, i.e., $r/L = \alpha = \text{const.}$ With this choice, the number of SLIOMs remains extensive, while its fraction relative to the total number of lattice sites can be controlled by $\alpha$:
\begin{equation}
Q = (2\rho - 2\alpha - 1)L.
\end{equation}
Note that the number of blocks, $B$, is exponential in $Q$, and thus also in $L$. Consequently, the Hilbert-space fragmentation occurs for any $\alpha < \rho - 1/2$. }

\textcolor{black}{It remains to be clarified how the rescaled fidelity susceptibility and autocorrelation function behave in the thermodynamic limit. We first focus on $\chi_\text{av}\mu$ and note that it is characterized by two parameters, i.e., the energy cutoff $\mu$ and the system size $L$. Importantly, the limits $L \to \infty$ and $\mu \to 0$ are not expected to commute, in the same way that the limits $L \to \infty$ and $t \to \infty$ are known not to. Since in the conventional experimental setups the system size is effectively infinite, it is natural to first take the limit $L \to \infty$. For a fixed energy cutoff $\mu$, the number of peaks increases with $L$, but each peak also becomes broader, and the larger the cutoff, the broader the peaks (see, for example, the discussion in Ref.~\cite{Kim_2024}). While we can only speculate about the limit $L \to \infty$, we expect that all peaks eventually overlap, forming a single broad peak that spans the entire extended critical regime. Following a similar line of thought for the autocorrelation function, we expect that its long-time value remains nonthermal in the thermodynamic limit provided that the density of SLIOMs is non-vanishing, i.e., $Q/L \ne 0$ for $L \to \infty$.}

\section{Conclusions}
In this paper, we have introduced a protocol that enables tuning of the Hilbert-space fragmentation in the $t$–$J_z$ model. \textcolor{black}{Specifically, by progressively including hopping terms, more particles can move through sites with $N_Z>2$, resulting in the successive breaking of conservation laws (SLIOMs).} We have derived analytical expressions for the numbers of SLIOMs and blocks. We have also performed direct numerical calculations, demonstrating that the existence of SLIOMs is reflected in spin dynamics, leading to an exceptionally broad range of relaxation times, a feature difficult to realize in integrable or nearly integrable models.

Finally, we have also studied the behavior of the rescaled fidelity susceptibility, which measures the sensitivity of energy eigenstates to modifications of the Hamiltonian. We have found that our model with tunable Hilbert-space fragmentation supports an extended critical regime, where the rescaled fidelity susceptibility exhibits multiple maxima. Each maximum signals a change in the number of SLIOMs, as well as an ultra-slow relaxation of local observables. Moreover, the number of maxima increases with the number of lattice sites. We emphasize that this is a highly nontrivial behavior, as typical integrability-breaking transitions feature a single peak signaling the breaking of all LIOMs.

\acknowledgements 
We acknowledge discussions with L.~Vidmar, A.~Polkovnikov, D.~Sels. and J. Paw{\l}owski. J.~B. acknowledges support from program No. P1-0044 of the Slovenian Research Agency (ARIS). M.M.
acknowledges support by the National Science Centre
(NCN), Poland via project 2020/37/B/ST3/00020. Numerical studies in this work have been carried out using resources provided by the~Wroclaw Centre for Networking and Supercomputing, Grant No. 579 (M. L., P.~{\L}.). The data that support the findings of this study are openly available~\cite{lisiecki_2025_17434301}.

\appendix

\section{Examples of tunable Hilbert-space fragmentation.} \label{app:a}
 
\textcolor{black}{The data presented in Tab.~\ref{tab:number_of_blocks} serve a twofold purpose. First, they illustrate how the number of blocks changes with the number of rungs for the systems studied in detail in the main text. Second, they show that the lower bound from Eq.~\eqref{bnumber} yields the exact number of blocks in the studied system, an agreement that we found to hold for all configurations of parameters considered.}

\begin{table}[htb!]
\centering
\begin{tabular}{|>{\centering\arraybackslash}p{0.3\columnwidth}|
                >{\centering\arraybackslash}p{0.09\columnwidth}|
                >{\centering\arraybackslash}p{0.09\columnwidth}|
                >{\centering\arraybackslash}p{0.09\columnwidth}|
                >{\centering\arraybackslash}p{0.09\columnwidth}|
                >{\centering\arraybackslash}p{0.09\columnwidth}|
                >{\centering\arraybackslash}p{0.09\columnwidth}|}
\hline
$r$ & 0 & 1 & 2 & 3 & 4 & 5 \\ \hline
$B\,(L=12)$ & 70 & 4 & 1 & 1 & 1 & 1 \\ \hline
$B\,(L=14)$ & 924 & 256 & 64 & 16 & 4 & 1 \\ \hline
$B\,(L=16)$ & 8192 & 256 & 64 & 16 & 4 & 1 \\ \hline
\end{tabular}
\caption{The number of blocks, $B$, versus the number of additional rungs, $r$, for different ladders with $L$ sites, $N_h$ holes, and total spin $S^z$. We consider $L=12$ with $N_h=4$ and $L=14$ with $N_h=2$. Both ladders have $S^z=0$. Additionally, we include $L=16$ with $N_h=3$ and $S^z=1/2$.}
\label{tab:number_of_blocks}
\end{table}

\section{Derivation of Eq.~(\ref{eq:Chi_final}).} \label{app:b}

The fidelity susceptibility naturally arises when considering the sensitivity of a system to changes in the Hamiltonian. For example, let us consider
\begin{equation}
    H(\lambda) = H_0 + \lambda O,
\end{equation}
and we refer to $\partial_\lambda H = O$ as a perturbation. The modification of energy eigenstates, $H(\lambda)|n(\lambda)\rangle = E(\lambda) |n(\lambda)\rangle$, can be quantified by the following overlap (up to the second order in $d\lambda$)~\cite{You_2007}
\begin{equation}
    |\langle n(\lambda)|n(\lambda+d\lambda)\rangle|^2 \approx 1 - \chi_n d\lambda^2,
\end{equation}
where the coefficient $\chi_n$ is known as the fidelity susceptibility and is given by
\begin{equation}
\label{eqA:Chi_n}
\chi_n = \sum_{m \neq n} \frac{|\langle n (\lambda) | O | m (\lambda) \rangle|^2}{\omega_{nm}^2}.
\end{equation}
In the above equation, we defined $\omega_{nm}=E_n(\lambda)-E_m(\lambda)$. From now on, we focus on the limit $\lambda\rightarrow 0$ and we make the dependence on $\lambda$ implicit.

Intuitively, ergodic systems are expected to be chaotic and more sensitive to perturbations than integrable ones. Therefore the fidelity susceptibility, when averaged over all energy eigenstates, can serve as an indicator of the integrability-breaking transitions~\cite{PhysRevX.10.041017}. To avoid an ill-defined denominator for degenerate energy eigenstates, and to probe the system's dynamics at a selected energy scale $\mu$ (or at selected timescale $\tau = \mu^{-1}$), we consider its regularized version
\begin{equation} 
\label{eqA:Chi_av} 
\chi_\text{av} = \frac{1}{Z}\sum_{n=1}^{Z}\sum_{m\neq n} \frac{\omega_{nm}^2}{(\omega_{nm}^2+\mu^2)^2}|\langle n | O | m \rangle|^2, 
\end{equation}
where $Z$ is the Hilbert space dimension (in the given symmetry sector). The energy cutoff is usually restricted to $\omega_{H}<\mu\ll\Delta E$, where $\omega_H\propto Z^{-1}$ is the Heisenberg energy and $\Delta E$ is the energy bandwidth. To simplify the expression from Eq.~\eqref{eqA:Chi_av}, we replace the sums with integrals and perform the substitution of variables, i.e., $E_n=E+\omega/2$ and $E_m=E-\omega/2$. We arrive at
\begin{equation}
     \chi_\text{av}=\frac{1}{Z}\int\limits_{-\infty}^{\infty}\int\limits_{-\infty}^{\infty}\mspace{-3mu}dEd\omega\mspace{2mu}\rho\mspace{-4mu}\left(E+\frac{\omega}{2}\right)\rho\mspace{-4mu}\left(E-\frac{\omega}{2}\right)\frac{\omega^2|\langle n | O | m \rangle|^2}{(\omega^2+\mu^2)^2},
\end{equation}
where $\rho(E)\propto Z$ is the density of states. Taking into account that 
\begin{itemize}
    \item offdiagonal matrix elements, both in ergodic~\cite{Srednicki_1999,Sch_nle_2021} and integrable~\cite{Brenes_2020b,Brenes_2020,PhysRevE.102.062113} systems, can be approximated as $|\langle n | O | m \rangle|^2 \approx \rho(E)^{-1} |f(E,\omega)|^2$, where $f(E,\omega)$ is a smooth function of its arguments,
    \item $\Delta(\omega)=(2\mu/\pi)/[\omega^2/(\omega^2+\mu^2)^2]$ is a probability distribution function sharply peaked at $\omega = \mu$,
    \item $\rho(E \pm \mu/2) \approx \rho(E)$,
\end{itemize}
we establish the relation from the main text:
\begin{equation}
\label{eqA:Chi_final}
    \chi_\text{av}\mu\propto\frac{1}{ Z}\int\limits_{-\infty}^{\infty}\mspace{-3mu}dE \rho(E) |f(E,\mu)|^2=|f(\mu)|^2.
\end{equation}
Finally, we highlight that the average over mean energies, $|f(\omega)|^2$, corresponds to the spectral function, which is the Fourier transform of the autocorrelation function, 
\begin{equation}
|f(\omega)|^2=(2\pi Z)^{-1} \sum_{n=1}^{Z}\int\limits_{-\infty}^{\infty} dt e^{i\omega t}[\langle n|O(t)O|n\rangle-|\langle n | O|n\rangle|^2]\,.
\end{equation}

\section{Details of numerical calculations.} \label{app:c}

Fidelity susceptibilities, $\chi_\text{av}$, were calculated as
\begin{equation}
    \chi_\text{av} = \frac{1}{D} \sum_{n=1}^{D} \sum_{\substack{m=1,\\ m \neq n}}^{Z} \frac{\omega_{nm}^2}{(\omega_{nm}^2 + \mu^2)^2} \left| \langle n | \partial_\lambda H | m \rangle \right|^2, 
\end{equation}
where $Z$ is the Hilbert space dimension (in the given symmetry sector). For the systems with $L=12,N=8,S^z=0$, the averaging was performed over $D=800$ energy eigenstates from the middle of the spectrum. For $L=14,N=12,S^z=0$, the averaging was performed over $D=4000$ ($D=800$) energy eigenstates from the middle of the spectrum when $r \le 3$ ($r > 3$). 

According to Eq.~\eqref{eq:Chi_final} from the main text, the behavior of the fidelity susceptibility, $\chi_\text{av}$, is encoded in the spectral function, $|f(\omega)|^2$. The latter function can be calculated from coarse-grained offdiagonal matrix elements:
\begin{equation}
    |f(\omega)|^2 \approx \frac{Z}{D} \sum_{\substack{ n, m\neq n \\ |(E_n + E_m)/2| < \Delta E \\ |(E_n - E_m) - \omega| < \Delta\omega }} |\langle n | \partial_\lambda H | m \rangle|^2,
\end{equation}
where $Z$ is the Hilbert space dimension (in the given symmetry sector), $\Delta E = 0.025L$ and $\Delta \omega$ was established by dividing the range $\log_{10}(\omega_H)<\log_{10}(\omega)<0$ into $160$ bins, where $\omega_H\approx 10^{-4}$ is the Heisenberg energy. Additionally, $D$ is the number of offdiagonal matrix elements that satisfy the conditions $|(E_n + E_m)/2| < \Delta E$ and $|(E_n - E_m) - \omega| < \Delta\omega$. In the final step, we applied a moving average over $20$ bins to smooth out the oscillations visible in the low-$\omega$ regime.

\begin{figure}[t!]
\includegraphics[width=\columnwidth]{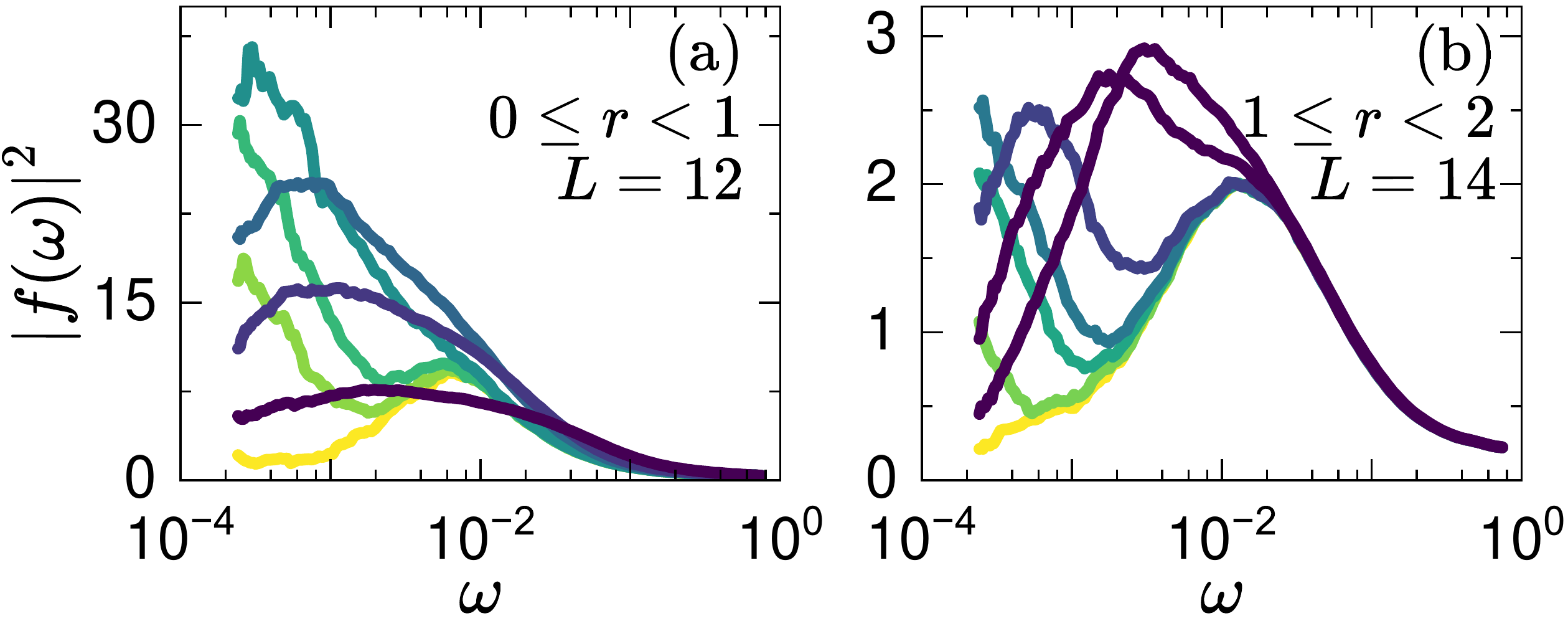}
\caption{(a)~Spectral functions $|f(\omega)|^2$ shown for a ladder with $L=12$ near the eigenstate transition at $r=0$, i.e., $r=0$, $0.01$, $0.02$, $0.05$, $0.2$, $0.3$ and $0.6$. (b)~Same results for $L=14$ and $r=1$, i.e., $r=1$, $1.004$, $1.01$, $1.015$, $1.03$, $1.1$ and $1.2$. Darker colors indicate larger $r$.}
\label{fig:Fig2_T2}
\end{figure}

We now show how the spectral function $|f(\omega)|^2$ evolves with $r$ for the perturbation studied in the main text, i.e., $\partial_\lambda H=\frac{1}{\sqrt{L}} \sum_{\sigma = \uparrow, \downarrow} \sum_{i=1}^{L-2} ( c_{i, \sigma}^\dagger c_{i+2, \sigma} + \text{H.c.})$. In panels (a) and (b) of Fig.~\ref{fig:Fig2_T2}, we present results for a ladder with $L=12,N=8,S^z=0$ near $r=0$ and $L=14,N=12,S^z=0$ near $r=1$, respectively. Darker colors indicate larger $\left\{r\right\}$. We observe that for $\left\{r\right\} \approx 0$, the spectral function $|f(\omega)|^2$ exhibits a gap, which is characteristic of integrable systems and has been related to the existence of LIOMs. In the fragmented system considered here, this feature is most likely associated with the presence of SLIOMs. As $\left\{r\right\}$ increases, $|f(\omega)|^2$ develops a polynomial tail, indicating an exceptionally slow relaxation of the autocorrelation function of $\partial_\lambda H$. This results from the breaking of a subset of SLIOMs. Eventually, as the relaxation times of the former SLIOMs decrease and become much shorter than the Heisenberg time $\tau_H = \omega_H^{-1}$, the polynomial tail of $|f(\omega)|^2$ shifts to higher $\omega$, revealing a gap associated with the remaining SLIOMs.

We emphasize that when calculating $|f(\omega)|^2$ and $\chi_\text{av}$, we add the term $0.5n_{L,\uparrow} + n_{L,\downarrow}$ to the Hamiltonian to break the parity and spin-inversion symmetries. It does not affect the Hilbert-space fragmentation and becomes relevant only in the ergodic regime, where the Hamiltonian comprises a single block.

\section{$\chi_\text{av}\mu$ and $|f(\omega)|^2$ for different $\partial_\lambda H$.} \label{app:d}

In the main part of this manuscript, we considered the next-nearest neighbor hopping as a perturbation, i.e., $\partial_\lambda H=T_2=\frac{1}{\sqrt{L}} \sum_{\sigma = \uparrow, \downarrow} \sum_{i=1}^{L-2} ( c_{i, \sigma}^\dagger c_{i+2, \sigma} + \text{H.c.})$. This choice was motivated by its locality and the presence of nonzero matrix elements between states from different blocks. It ensures that the behavior of the rescaled fidelity susceptibility, $\chi_\text{av}\mu$, and the spectral function, $|f(\omega)|^2$, is not a trivial consequence of $\partial_\lambda H$ sharing the block structure with $H$ at $\lambda=0$. Here, we extend the study to two additional perturbations. Particularly, the site occupation in the middle of the ladder,
\begin{equation}
    \partial_\lambda H = n_{L/2} =\sum_{\sigma=\uparrow,\downarrow} c_{L/2,\sigma}^\dagger c_{L/2,\sigma},
\end{equation}
which is local and diagonal in the computational basis, and the occupation of the zero-momentum mode, 
\begin{equation}
\partial_\lambda H = m_0 = \sum_{\sigma=\uparrow,\downarrow}\sum_{i,j=1}^{L} c_{i,\sigma}^\dagger c_{j,\sigma},
\end{equation}
which is nonlocal (but few-body) and couples states from different blocks.

\begin{figure}[t!]
\includegraphics[width=\columnwidth]{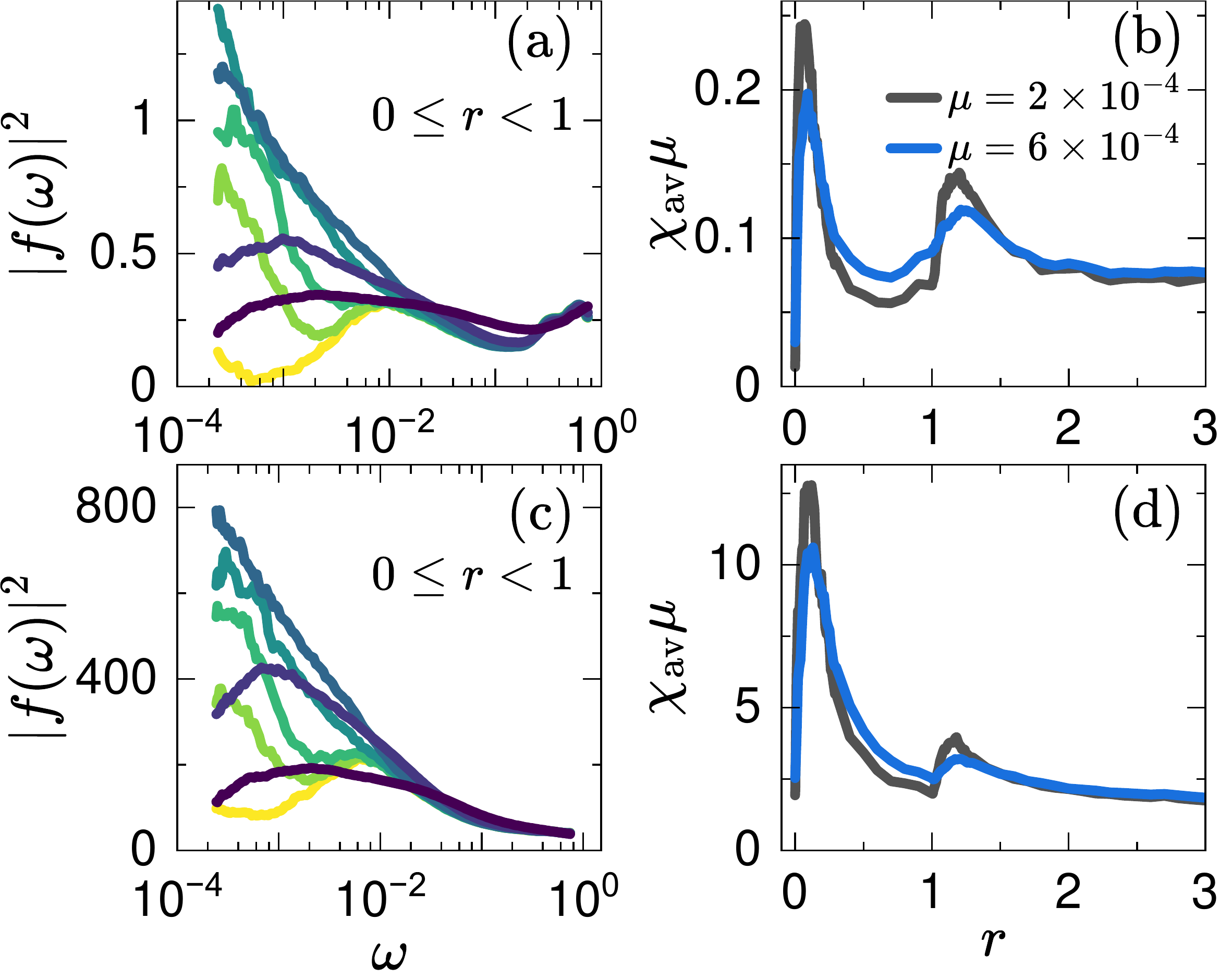}
\caption{Numerical results for a ladder with $L=12$. Spectral functions $|f(\omega)|^2$ near the eigenstate transition at $r=0$ for (a)~$n_{L/2}$ and (c)~$m_0$. We consider $r=0$, $0.01$, $0.02$, $0.05$, $0.1$, $0.25$, $0.6$ and darker colors indicate larger $r$. Rescaled fidelity susceptibilities $\chi_\text{av}\mu$ plotted against $r$ for (b)~$n_{L/2}$ and (d)~$m_0$. We consider $\mu = 2 \times 10^{-4}$ and $6 \times 10^{-4}$.}
\label{fig_n_m0_L12}
\end{figure}

\begin{figure}[t!]
\includegraphics[width=\columnwidth]{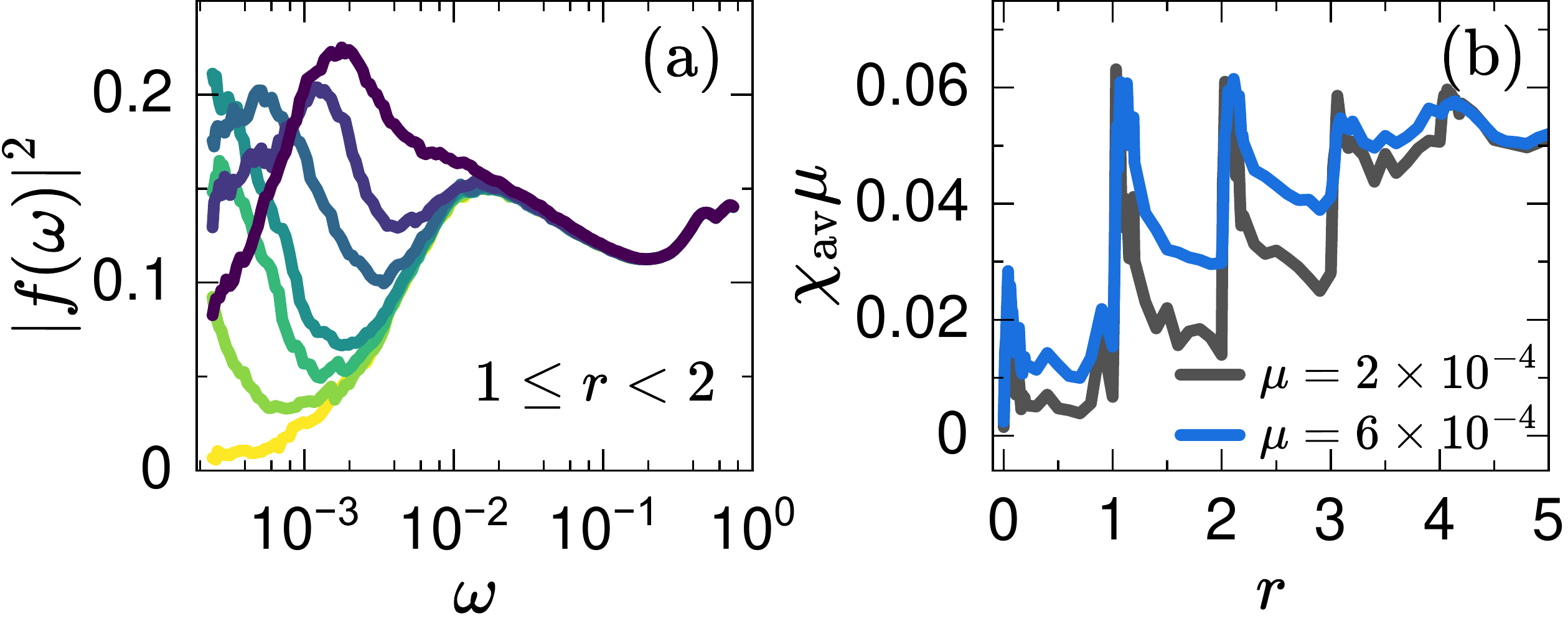}
\caption{Numerical results for a ladder with $L=14$. We focus on the perturbation $n_{L/2}$. (a)~Spectral functions $|f(\omega)|^2$ near the eigenstate transition at $r=1$. We consider $r=1$, $1.004$, $1.01$, $1.015$, $1.03$, $1.05$, $1.1$ and darker colors indicate larger $r$. (b)~Rescaled fidelity susceptibilities $\chi_\text{av}\mu$ plotted against $r$. We consider $\mu = 2 \times 10^{-4}$ and $6 \times 10^{-4}$.}
\label{fig_n_L14}
\end{figure}

Results of numerical calculations for a ladder with $L=12,N=8,S^z=0$ are shown in Fig.~\ref{fig_n_m0_L12}. The spectral functions $|f(\omega)|^2$ near the eigenstate transition at $r=0$ are presented in panels (a) and (c) for the site occupation $n_{L/2}$ and the momentum mode occupation $m_0$, respectively. Darker colors indicate larger $r$. Additionally, the rescaled fidelity susceptibilities $\chi_\text{av}\mu$ are plotted against $r$ in panel (b) for $n_{L/2}$ and in panel (d) for $m_0$. We consider $\mu = 2 \times 10^{-4}$ and $6 \times 10^{-4}$. Results of numerical calculations for a system with $L=14,N=12,S^z=0$ are shown in Fig.~\ref{fig_n_L14}. Focusing on $n_{L/2}$, the spectral functions $|f(\omega)|^2$ near the eigenstate transition at $r=1$ are depicted in panel (a), while the rescaled fidelity susceptibility $\chi_\text{av}\mu$ is plotted against $r$ in panel (b). Clearly, $|f(\omega)|^2$ and $\chi_\text{av}\mu$ exhibit qualitatively similar behavior for all considered perturbations.

We complement the analysis by showing that the eigenstate transitions can also be inferred from the distributions of offdiagonal matrix elements. Deep in the ergodic regime, these distributions are expected to be Gaussian, at least for energy eigenstates with close energies~\cite{PhysRevE.87.012118,Brenes_2020,Brenes_2020b}. Therefore we consider the measure~\cite{PhysRevE.100.062134},
\begin{equation}
    \Gamma=\frac{\overline{|\langle n |\partial_\lambda H|m\rangle|^2}}{\overline{|\langle n |\partial_\lambda H|m\rangle|}^2}
\end{equation}
with
\begin{equation}
    \overline{f(\langle n |O| m \rangle)}=\sum_{\substack{n,m\neq n \\ |(E_n + E_m)/2| < 0.025L \\ |E_n-E_m|<0.02}} f(\langle n |O| m \rangle),
\end{equation}
which equals $\pi/2$ for a Gaussian distribution.

\begin{figure}[t!]
\includegraphics[width=\columnwidth]{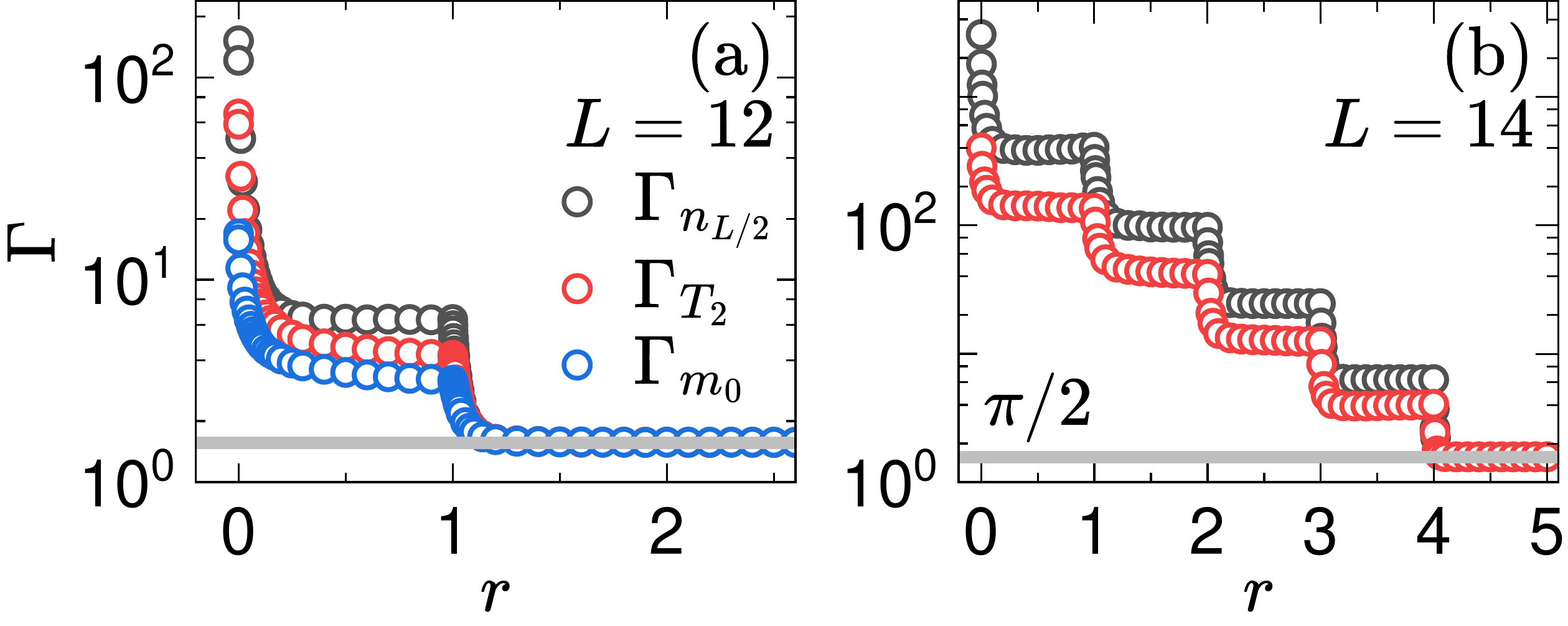}
\caption{The measure $\Gamma$ plotted against $r$ for a ladder with (a) $L=12$ and (b) $L=14$. We consider three perturbations, i.e., the next-nearest neighbor hopping $T_2$, the site occupation $n_{L/2}$, and the zero momentum mode occupation $m_0$ (for $L=12$). The horizontal lines mark the expectation for the Gaussian distribution, i.e., $\Gamma=\pi/2$.}
\label{fig_distrib}
\end{figure}

In Fig.~\ref{fig_distrib}, we present $\Gamma$ as functions of $r$ for all considered $\partial_\lambda H$. Results for ladders with $L=12,N=8,S^z=0$ and $L=14,N=12,S^z=0$ are shown in panels (a) and (b), respectively. We first note that, in agreement with expectations, $\Gamma \approx \pi/2$ when the Hamiltonian consists of a single block. This occurs for $r > 1$ when $L = 12$ and for $r > 4$ when $L = 14$. Moreover, $\Gamma$ exhibits a jump near each eigenstate transition and remains approximately constant between the transitions. This behavior is the most easily explained for the site occupation, $\partial_\lambda H = n_{L/2}$, as it is diagonal in the computational basis. Since its offdiagonal matrix elements between energy eigenstates from different blocks are exactly zero, their number increases with decreasing $r$, leading to the emergence of a singularity in the distribution. Similar behavior has been observed for local observables when the system approaches the integrable point, i.e., the distribution of offdiagonal matrix elements develops a peak around zero so that it can be modeled as a combination of two Gaussians~\cite{PhysRevE.91.012144}.

\section{Scaling of the maxima of $\chi_\text{av}$.} \label{app:e}

In this section, we study the behavior of $\chi_\text{av}$ with $\mu^{-1}$ in detail. We recall that $\chi_\text{av}$ is expected to scale as $\mu^{-\alpha}$ with $\alpha \in (1,2]$ in the vicinity of its maxima~\cite{lim2024,Kim_2024}. (Note that in the main text, we study $\chi_\text{av} \mu$ instead of $\chi_\text{av}$, so that $\alpha \in (0,1]$ rather than $(1,2]$.) Simultaneously, it scales as $\mu^{-1}$ in ergodic systems, while it increases more slowly than polynomially in integrable and generally nonergodic systems.

\begin{figure}[b!]
\includegraphics[width=\columnwidth]{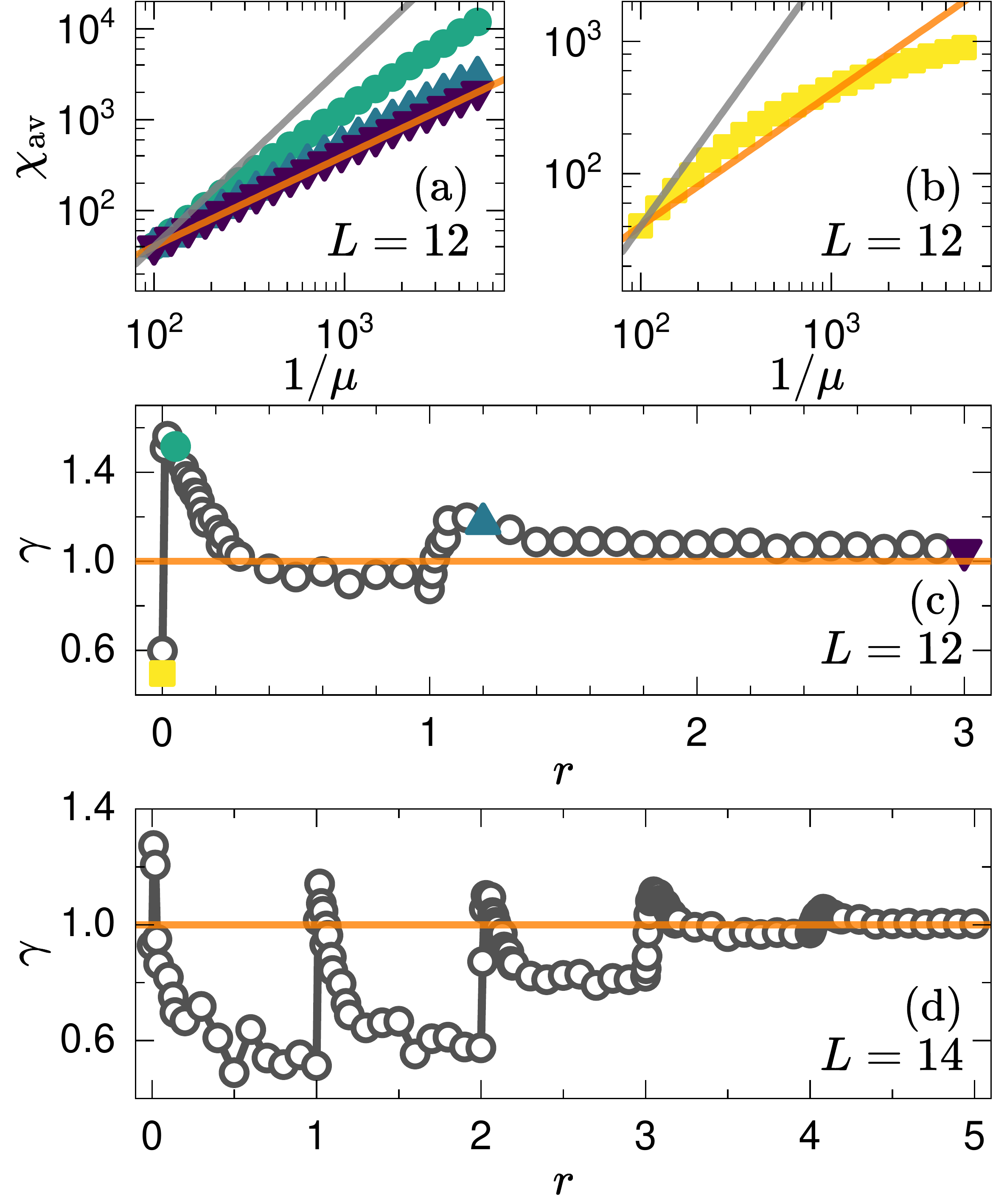}
\caption{[(a) and (b)] The fidelity susceptibility $\chi_\text{av}$ plotted against the inverse of energy cutoff $\mu^{-1}$ for a ladder with $L=12$. In (a), we consider $r=0.05$ (circles), $1.2$ (upward triangles), $3$ (downward triangles). In (b), we consider $r=0$ (squares). Two solid lines mark the limiting predictions, i.e., $\chi_\text{av}\propto \mu^{-1}$ and $\chi_\text{av}\propto \mu^{-2}$. The curves from (a) and (b) correspond to the highlighted points from (c). The derivative $\gamma$ as a function of $r$ for a ladder with (c) $L=12$ and (d) $14$. The solid lines mark the prediction for ergodic systems, i.e., $\gamma=1$.}
\label{fig_chi_derivative_T2_L12}
\end{figure}

In Figs.~\ref{fig_chi_derivative_T2_L12}(a) and~\ref{fig_chi_derivative_T2_L12}(b), we plot $\chi_\text{av}$ against $\mu^{-1}$ for selected $r$. We focus on the perturbation from the main text, i.e., $\partial_\lambda H=\frac{1}{\sqrt{L}} \sum_{\sigma = \uparrow, \downarrow} \sum_{i=1}^{L-2} ( c_{i, \sigma}^\dagger c_{i+2, \sigma} + \text{H.c.})$, and we consider a ladder with $L=12,N=8,S^z=0$. As indicated in panel (a) by circles an upward triangles, the behavior of the fidelity susceptibility near the eigenstate transitions is consistent with the scaling $\mu^{-\alpha}$ with $\alpha \in (1,2]$. As indicated in the same panel by downward triangles, $\chi_\text{av}$ in the non-fragmented system closely follows $\mu^{-1}$. Finally, the fidelity susceptibility calculated in the fragmented system far from eigenstate transitions shows bending in the log-log scale, suggesting that $\chi_\text{av}$ increases more slowly than polynomially with $\mu^{-1}$, see panel (b). The curves from Figs.~\ref{fig_chi_derivative_T2_L12}(a) and~\ref{fig_chi_derivative_T2_L12}(b) correspond to the highlighted points in Fig.~\ref{fig_chi_derivative_T2_L12}(c).

We complement the study with the calculation of
\begin{equation} 
    \gamma = \frac{\partial \log_{10} \chi_\text{av}}{\partial \log_{10} \mu^{-1}}.
\end{equation} 
Numerically, $\gamma$ is determined by calculating few two-point derivatives in the range $\mu \in [2 \times 10^{-4}, 12 \times 10^{-4}]$ and averaging the results. We note that if $\chi_\text{av} \propto \mu^{-\alpha}$ then $\gamma = \alpha$. In Figs.~\ref{fig_chi_derivative_T2_L12}(c) and~\ref{fig_chi_derivative_T2_L12}(d), we plot $\gamma$ against $r$ for a ladders with $L=12,N=8,S^z=0$ and $L=14,N=12,S^z=0$, respectively. The solid lines mark the prediction for ergodic systems, i.e., $\gamma=1$. In agreement with previous findings, we observe $\gamma < 1$ for fragmented systems, $\gamma = 1$ for non-fragmented systems, and $\gamma > 1$ near eigenstate transitions.

\FloatBarrier
\bibliographystyle{biblev1}
\bibliography{references}

\begin{thebibliography}{10}
\expandafter\ifx\csname url\endcsname\relax
  \def\url#1{{\tt #1}}\fi
\expandafter\ifx\csname urlprefix\endcsname\relax\def\urlprefix{URL }\fi
\expandafter\ifx\csname bibinfo\endcsname\relax\def\bibinfo#1#2{#2}\fi
\expandafter\ifx\csname eprint\endcsname\relax\def\eprint#1{\url{#1}}\fi

\bibitem{Rigol2008}
\bibinfo{author}{M.~Rigol}, \bibinfo{author}{V.~Dunjko}, and \bibinfo{author}{M.~Olshanii}, \bibinfo{title}{{Thermalization and its mechanism for generic isolated quantum systems}}, \bibinfo{journal}{\href{http://dx.doi.org/10.1038/nature06838}{Nature}} \href{http://dx.doi.org/10.1038/nature06838}{{\bf \bibinfo{volume}{452}}, \bibinfo{pages}{854}}  (\href{http://dx.doi.org/10.1038/nature06838}{\bibinfo{year}{2008}}).

\bibitem{Eisert_2015}
\bibinfo{author}{J.~Eisert}, \bibinfo{author}{M.~Friesdorf}, and \bibinfo{author}{C.~Gogolin}, \bibinfo{title}{{Quantum many-body systems out of equilibrium}}, \bibinfo{journal}{\href{http://dx.doi.org/10.1038/nphys3215}{Nat. Phys.}} \href{http://dx.doi.org/10.1038/nphys3215}{{\bf \bibinfo{volume}{11}}, \bibinfo{pages}{124}}  (\href{http://dx.doi.org/10.1038/nphys3215}{\bibinfo{year}{2015}}).

\bibitem{Deutsch_2018}
\bibinfo{author}{J.~M. Deutsch}, \bibinfo{title}{{Eigenstate thermalization hypothesis}}, \bibinfo{journal}{\href{http://dx.doi.org/10.1088/1361-6633/aac9f1}{Rep. Prog. Phys.}} \href{http://dx.doi.org/10.1088/1361-6633/aac9f1}{{\bf \bibinfo{volume}{81}}, \bibinfo{pages}{082001}}  (\href{http://dx.doi.org/10.1088/1361-6633/aac9f1}{\bibinfo{year}{2018}}).

\bibitem{Sierant_2019b}
\bibinfo{author}{P.~Sierant}, \bibinfo{author}{A.~Maksymov}, \bibinfo{author}{M.~Kuś}, and \bibinfo{author}{J.~Zakrzewski}, \bibinfo{title}{{Fidelity susceptibility in Gaussian random ensembles}}, \bibinfo{journal}{\href{http://dx.doi.org/10.1103/physreve.99.050102}{Phys. Rev. E}} \href{http://dx.doi.org/10.1103/physreve.99.050102}{{\bf \bibinfo{volume}{99}}}  (\href{http://dx.doi.org/10.1103/physreve.99.050102}{\bibinfo{year}{2019}}).

\bibitem{PhysRevX.10.041017}
\bibinfo{author}{M.~Pandey}, \bibinfo{author}{P.~W. Claeys}, \bibinfo{author}{D.~K. Campbell}, \bibinfo{author}{A.~Polkovnikov}, and \bibinfo{author}{D.~Sels}, \bibinfo{title}{{Adiabatic Eigenstate Deformations as a Sensitive Probe for Quantum Chaos}}, \bibinfo{journal}{\href{http://dx.doi.org/10.1103/PhysRevX.10.041017}{Phys. Rev. X}} \href{http://dx.doi.org/10.1103/PhysRevX.10.041017}{{\bf \bibinfo{volume}{10}}, \bibinfo{pages}{041017}}  (\href{http://dx.doi.org/10.1103/PhysRevX.10.041017}{\bibinfo{year}{2020}}).

\bibitem{lim2024}
\bibinfo{author}{C.~Lim}, \bibinfo{author}{K.~Matirko}, \bibinfo{author}{H.~Kim}, \bibinfo{author}{A.~Polkovnikov}, and \bibinfo{author}{M.~O. Flynn}, \href{https://arxiv.org/abs/2401.01927}{\bibinfo{title}{{Defining classical and quantum chaos through adiabatic transformations}}}  (\bibinfo{year}{2024}). \eprint{2401.01927}.

\bibitem{Mukhin2009}
\bibinfo{author}{E.~Mukhin}, \bibinfo{author}{V.~Tarasov}, and \bibinfo{author}{A.~Varchenko}, \bibinfo{title}{{Bethe Algebra of Homogeneous XXX Heisenberg Model has Simple Spectrum}}, \bibinfo{journal}{\href{http://dx.doi.org/10.1007/s00220-009-0733-4}{Commun. Math. Phys.}} \href{http://dx.doi.org/10.1007/s00220-009-0733-4}{{\bf \bibinfo{volume}{288}}, \bibinfo{pages}{1}}  (\href{http://dx.doi.org/10.1007/s00220-009-0733-4}{\bibinfo{year}{2009}}).

\bibitem{Kirillov_2014}
\bibinfo{author}{A.~N. Kirillov} and \bibinfo{author}{R.~Sakamoto}, \bibinfo{title}{{Singular solutions to the Bethe ansatz equations and rigged configurations}}, \bibinfo{journal}{\href{http://dx.doi.org/10.1088/1751-8113/47/20/205207}{J. Phys. A: Math. Theor.}} \href{http://dx.doi.org/10.1088/1751-8113/47/20/205207}{{\bf \bibinfo{volume}{47}}, \bibinfo{pages}{205207}}  (\href{http://dx.doi.org/10.1088/1751-8113/47/20/205207}{\bibinfo{year}{2014}}).

\bibitem{PhysRevLett.98.050405}
\bibinfo{author}{M.~Rigol}, \bibinfo{author}{V.~Dunjko}, \bibinfo{author}{V.~Yurovsky}, and \bibinfo{author}{M.~Olshanii}, \bibinfo{title}{{Relaxation in a Completely Integrable Many-Body Quantum System: An Ab Initio Study of the Dynamics of the Highly Excited States of 1D Lattice Hard-Core Bosons}}, \bibinfo{journal}{\href{http://dx.doi.org/10.1103/PhysRevLett.98.050405}{Phys. Rev. Lett.}} \href{http://dx.doi.org/10.1103/PhysRevLett.98.050405}{{\bf \bibinfo{volume}{98}}, \bibinfo{pages}{050405}}  (\href{http://dx.doi.org/10.1103/PhysRevLett.98.050405}{\bibinfo{year}{2007}}).

\bibitem{PhysRevA.74.053616}
\bibinfo{author}{M.~Rigol}, \bibinfo{author}{A.~Muramatsu}, and \bibinfo{author}{M.~Olshanii}, \bibinfo{title}{{Hard-core bosons on optical superlattices: Dynamics and relaxation in the superfluid and insulating regimes}}, \bibinfo{journal}{\href{http://dx.doi.org/10.1103/PhysRevA.74.053616}{Phys. Rev. A}} \href{http://dx.doi.org/10.1103/PhysRevA.74.053616}{{\bf \bibinfo{volume}{74}}, \bibinfo{pages}{053616}}  (\href{http://dx.doi.org/10.1103/PhysRevA.74.053616}{\bibinfo{year}{2006}}).

\bibitem{Vidmar_2016}
\bibinfo{author}{L.~Vidmar} and \bibinfo{author}{M.~Rigol}, \bibinfo{title}{{Generalized Gibbs ensemble in integrable lattice models}}, \bibinfo{journal}{\href{http://dx.doi.org/10.1088/1742-5468/2016/06/064007}{J. Stat. Mech.: Theory Exp.}} \href{http://dx.doi.org/10.1088/1742-5468/2016/06/064007}{{\bf \bibinfo{volume}{2016}}, \bibinfo{pages}{064007}}  (\href{http://dx.doi.org/10.1088/1742-5468/2016/06/064007}{\bibinfo{year}{2016}}).

\bibitem{PhysRevB.107.184312}
\bibinfo{author}{P.~Orlov}, \bibinfo{author}{A.~Tiutiakina}, \bibinfo{author}{R.~Sharipov}, \bibinfo{author}{E.~Petrova}, \bibinfo{author}{V.~Gritsev}, and \bibinfo{author}{D.~V. Kurlov}, \bibinfo{title}{{Adiabatic eigenstate deformations and weak integrability breaking of Heisenberg chain}}, \bibinfo{journal}{\href{http://dx.doi.org/10.1103/PhysRevB.107.184312}{Phys. Rev. B}} \href{http://dx.doi.org/10.1103/PhysRevB.107.184312}{{\bf \bibinfo{volume}{107}}, \bibinfo{pages}{184312}}  (\href{http://dx.doi.org/10.1103/PhysRevB.107.184312}{\bibinfo{year}{2023}}).

\bibitem{Pozsgay_2024}
\bibinfo{author}{B.~Pozsgay}, \bibinfo{author}{R.~Sharipov}, \bibinfo{author}{A.~Tiutiakina}, and \bibinfo{author}{I.~Vona}, \bibinfo{title}{{Adiabatic gauge potential and integrability breaking with free fermions}}, \bibinfo{journal}{\href{http://dx.doi.org/10.21468/scipostphys.17.3.075}{SciPost Phys.}} \href{http://dx.doi.org/10.21468/scipostphys.17.3.075}{{\bf \bibinfo{volume}{17}}, \bibinfo{pages}{075}}  (\href{http://dx.doi.org/10.21468/scipostphys.17.3.075}{\bibinfo{year}{2024}}).

\bibitem{Mori_2018}
\bibinfo{author}{T.~Mori}, \bibinfo{author}{T.~N. Ikeda}, \bibinfo{author}{E.~Kaminishi}, and \bibinfo{author}{M.~Ueda}, \bibinfo{title}{{Thermalization and prethermalization in isolated quantum systems: a theoretical overview}}, \bibinfo{journal}{\href{http://dx.doi.org/10.1088/1361-6455/aabcdf}{J. Phys. B: At. Mol. Opt. Phys.}} \href{http://dx.doi.org/10.1088/1361-6455/aabcdf}{{\bf \bibinfo{volume}{51}}, \bibinfo{pages}{112001}}  (\href{http://dx.doi.org/10.1088/1361-6455/aabcdf}{\bibinfo{year}{2018}}).

\bibitem{PhysRevB.84.054304}
\bibinfo{author}{M.~Kollar}, \bibinfo{author}{F.~A. Wolf}, and \bibinfo{author}{M.~Eckstein}, \bibinfo{title}{{Generalized Gibbs ensemble prediction of prethermalization plateaus and their relation to nonthermal steady states in integrable systems}}, \bibinfo{journal}{\href{http://dx.doi.org/10.1103/PhysRevB.84.054304}{Phys. Rev. B}} \href{http://dx.doi.org/10.1103/PhysRevB.84.054304}{{\bf \bibinfo{volume}{84}}, \bibinfo{pages}{054304}}  (\href{http://dx.doi.org/10.1103/PhysRevB.84.054304}{\bibinfo{year}{2011}}).

\bibitem{PhysRevLett.115.180601}
\bibinfo{author}{B.~Bertini}, \bibinfo{author}{F.~H.~L. Essler}, \bibinfo{author}{S.~Groha}, and \bibinfo{author}{N.~J. Robinson}, \bibinfo{title}{{Prethermalization and Thermalization in Models with Weak Integrability Breaking}}, \bibinfo{journal}{\href{http://dx.doi.org/10.1103/PhysRevLett.115.180601}{Phys. Rev. Lett.}} \href{http://dx.doi.org/10.1103/PhysRevLett.115.180601}{{\bf \bibinfo{volume}{115}}, \bibinfo{pages}{180601}}  (\href{http://dx.doi.org/10.1103/PhysRevLett.115.180601}{\bibinfo{year}{2015}}).

\bibitem{PhysRevX.9.021027}
\bibinfo{author}{K.~Mallayya}, \bibinfo{author}{M.~Rigol}, and \bibinfo{author}{W.~De~Roeck}, \bibinfo{title}{{Prethermalization and Thermalization in Isolated Quantum Systems}}, \bibinfo{journal}{\href{http://dx.doi.org/10.1103/PhysRevX.9.021027}{Phys. Rev. X}} \href{http://dx.doi.org/10.1103/PhysRevX.9.021027}{{\bf \bibinfo{volume}{9}}, \bibinfo{pages}{021027}}  (\href{http://dx.doi.org/10.1103/PhysRevX.9.021027}{\bibinfo{year}{2019}}).

\bibitem{PhysRevB.109.L161109}
\bibinfo{author}{J.~Paw\l{}owski}, \bibinfo{author}{M.~Panfil}, \bibinfo{author}{J.~Herbrych}, and \bibinfo{author}{M.~Mierzejewski}, \bibinfo{title}{{Long-living prethermalization in nearly integrable spin ladders}}, \bibinfo{journal}{\href{http://dx.doi.org/10.1103/PhysRevB.109.L161109}{Phys. Rev. B}} \href{http://dx.doi.org/10.1103/PhysRevB.109.L161109}{{\bf \bibinfo{volume}{109}}, \bibinfo{pages}{L161109}}  (\href{http://dx.doi.org/10.1103/PhysRevB.109.L161109}{\bibinfo{year}{2024}}).

\bibitem{PhysRevLett.96.067202}
\bibinfo{author}{P.~Jung}, \bibinfo{author}{R.~W. Helmes}, and \bibinfo{author}{A.~Rosch}, \bibinfo{title}{{Transport in Almost Integrable Models: Perturbed Heisenberg Chains}}, \bibinfo{journal}{\href{http://dx.doi.org/10.1103/PhysRevLett.96.067202}{Phys. Rev. Lett.}} \href{http://dx.doi.org/10.1103/PhysRevLett.96.067202}{{\bf \bibinfo{volume}{96}}, \bibinfo{pages}{067202}}  (\href{http://dx.doi.org/10.1103/PhysRevLett.96.067202}{\bibinfo{year}{2006}}).

\bibitem{PhysRevA.99.042117}
\bibinfo{author}{B.-B. Wei}, \bibinfo{title}{{Fidelity susceptibility in one-dimensional disordered lattice models}}, \bibinfo{journal}{\href{http://dx.doi.org/10.1103/PhysRevA.99.042117}{Phys. Rev. A}} \href{http://dx.doi.org/10.1103/PhysRevA.99.042117}{{\bf \bibinfo{volume}{99}}, \bibinfo{pages}{042117}}  (\href{http://dx.doi.org/10.1103/PhysRevA.99.042117}{\bibinfo{year}{2019}}).

\bibitem{PhysRevX.5.031007}
\bibinfo{author}{L.~Wang}, \bibinfo{author}{Y.-H. Liu}, \bibinfo{author}{J.~Imri\ifmmode~\check{s}\else \v{s}\fi{}ka}, \bibinfo{author}{P.~N. Ma}, and \bibinfo{author}{M.~Troyer}, \bibinfo{title}{{Fidelity Susceptibility Made Simple: A Unified Quantum Monte Carlo Approach}}, \bibinfo{journal}{\href{http://dx.doi.org/10.1103/PhysRevX.5.031007}{Phys. Rev. X}} \href{http://dx.doi.org/10.1103/PhysRevX.5.031007}{{\bf \bibinfo{volume}{5}}, \bibinfo{pages}{031007}}  (\href{http://dx.doi.org/10.1103/PhysRevX.5.031007}{\bibinfo{year}{2015}}).

\bibitem{PhysRevLett.103.170501}
\bibinfo{author}{D.~Schwandt}, \bibinfo{author}{F.~Alet}, and \bibinfo{author}{S.~Capponi}, \bibinfo{title}{{Quantum Monte Carlo Simulations of Fidelity at Magnetic Quantum Phase Transitions}}, \bibinfo{journal}{\href{http://dx.doi.org/10.1103/PhysRevLett.103.170501}{Phys. Rev. Lett.}} \href{http://dx.doi.org/10.1103/PhysRevLett.103.170501}{{\bf \bibinfo{volume}{103}}, \bibinfo{pages}{170501}}  (\href{http://dx.doi.org/10.1103/PhysRevLett.103.170501}{\bibinfo{year}{2009}}).

\bibitem{PhysRevB.77.245109}
\bibinfo{author}{S.-J. Gu}, \bibinfo{author}{H.-M. Kwok}, \bibinfo{author}{W.-Q. Ning}, and \bibinfo{author}{H.-Q. Lin}, \bibinfo{title}{{Fidelity susceptibility, scaling, and universality in quantum critical phenomena}}, \bibinfo{journal}{\href{http://dx.doi.org/10.1103/PhysRevB.77.245109}{Phys. Rev. B}} \href{http://dx.doi.org/10.1103/PhysRevB.77.245109}{{\bf \bibinfo{volume}{77}}, \bibinfo{pages}{245109}}  (\href{http://dx.doi.org/10.1103/PhysRevB.77.245109}{\bibinfo{year}{2008}}).

\bibitem{PhysRevA.77.032111}
\bibinfo{author}{S.~Chen}, \bibinfo{author}{L.~Wang}, \bibinfo{author}{Y.~Hao}, and \bibinfo{author}{Y.~Wang}, \bibinfo{title}{{Intrinsic relation between ground-state fidelity and the characterization of a quantum phase transition}}, \bibinfo{journal}{\href{http://dx.doi.org/10.1103/PhysRevA.77.032111}{Phys. Rev. A}} \href{http://dx.doi.org/10.1103/PhysRevA.77.032111}{{\bf \bibinfo{volume}{77}}, \bibinfo{pages}{032111}}  (\href{http://dx.doi.org/10.1103/PhysRevA.77.032111}{\bibinfo{year}{2008}}).

\bibitem{PhysRevLett.98.110601}
\bibinfo{author}{P.~Buonsante} and \bibinfo{author}{A.~Vezzani}, \bibinfo{title}{{Ground-State Fidelity and Bipartite Entanglement in the Bose-Hubbard Model}}, \bibinfo{journal}{\href{http://dx.doi.org/10.1103/PhysRevLett.98.110601}{Phys. Rev. Lett.}} \href{http://dx.doi.org/10.1103/PhysRevLett.98.110601}{{\bf \bibinfo{volume}{98}}, \bibinfo{pages}{110601}}  (\href{http://dx.doi.org/10.1103/PhysRevLett.98.110601}{\bibinfo{year}{2007}}).

\bibitem{PhysRevB.75.014439}
\bibinfo{author}{M.~Cozzini}, \bibinfo{author}{P.~Giorda}, and \bibinfo{author}{P.~Zanardi}, \bibinfo{title}{{Quantum phase transitions and quantum fidelity in free fermion graphs}}, \bibinfo{journal}{\href{http://dx.doi.org/10.1103/PhysRevB.75.014439}{Phys. Rev. B}} \href{http://dx.doi.org/10.1103/PhysRevB.75.014439}{{\bf \bibinfo{volume}{75}}, \bibinfo{pages}{014439}}  (\href{http://dx.doi.org/10.1103/PhysRevB.75.014439}{\bibinfo{year}{2007}}).

\bibitem{LeBlond_2021}
\bibinfo{author}{T.~LeBlond}, \bibinfo{author}{D.~Sels}, \bibinfo{author}{A.~Polkovnikov}, and \bibinfo{author}{M.~Rigol}, \bibinfo{title}{{Universality in the onset of quantum chaos in many-body systems}}, \bibinfo{journal}{\href{http://dx.doi.org/10.1103/physrevb.104.l201117}{Phys. Rev. B}} \href{http://dx.doi.org/10.1103/physrevb.104.l201117}{{\bf \bibinfo{volume}{104}}}  (\href{http://dx.doi.org/10.1103/physrevb.104.l201117}{\bibinfo{year}{2021}}).

\bibitem{Sels_2021}
\bibinfo{author}{D.~Sels} and \bibinfo{author}{A.~Polkovnikov}, \bibinfo{title}{{Dynamical obstruction to localization in a disordered spin chain}}, \bibinfo{journal}{\href{http://dx.doi.org/10.1103/physreve.104.054105}{Phys. Rev. E}} \href{http://dx.doi.org/10.1103/physreve.104.054105}{{\bf \bibinfo{volume}{104}}}  (\href{http://dx.doi.org/10.1103/physreve.104.054105}{\bibinfo{year}{2021}}).

\bibitem{świętek2025}
\bibinfo{author}{R.~\ifmmode \acute{S}\else \'{S}\fi{}wi\ifmmode~\mbox{\c{e}}\else \c{e}\fi{}tek}, \bibinfo{author}{P.~\L{}yd\ifmmode~\dot{z}\else \.{z}\fi{}ba}, and \bibinfo{author}{L.~Vidmar}, \bibinfo{title}{Fading ergodicity meets maximal chaos}, \bibinfo{journal}{\href{http://dx.doi.org/10.1103/PhysRevB.111.184203}{Phys. Rev. B}} \href{http://dx.doi.org/10.1103/PhysRevB.111.184203}{{\bf \bibinfo{volume}{111}}, \bibinfo{pages}{184203}}  (\href{http://dx.doi.org/10.1103/PhysRevB.111.184203}{\bibinfo{year}{2025}}).

\bibitem{PhysRevX.10.011047}
\bibinfo{author}{P.~Sala}, \bibinfo{author}{T.~Rakovszky}, \bibinfo{author}{R.~Verresen}, \bibinfo{author}{M.~Knap}, and \bibinfo{author}{F.~Pollmann}, \bibinfo{title}{{Ergodicity Breaking Arising from Hilbert Space Fragmentation in Dipole-Conserving Hamiltonians}}, \bibinfo{journal}{\href{http://dx.doi.org/10.1103/PhysRevX.10.011047}{Phys. Rev. X}} \href{http://dx.doi.org/10.1103/PhysRevX.10.011047}{{\bf \bibinfo{volume}{10}}, \bibinfo{pages}{011047}}  (\href{http://dx.doi.org/10.1103/PhysRevX.10.011047}{\bibinfo{year}{2020}}).

\bibitem{10.21468/SciPostPhys.15.3.093}
\bibinfo{author}{P.~Brighi}, \bibinfo{author}{M.~Ljubotina}, and \bibinfo{author}{M.~Serbyn}, \bibinfo{title}{{Hilbert space fragmentation and slow dynamics in particle-conserving quantum East models}}, \bibinfo{journal}{\href{http://dx.doi.org/10.21468/SciPostPhys.15.3.093}{SciPost Phys.}} \href{http://dx.doi.org/10.21468/SciPostPhys.15.3.093}{{\bf \bibinfo{volume}{15}}, \bibinfo{pages}{093}}  (\href{http://dx.doi.org/10.21468/SciPostPhys.15.3.093}{\bibinfo{year}{2023}}).

\bibitem{PhysRevB.101.174204}
\bibinfo{author}{V.~Khemani}, \bibinfo{author}{M.~Hermele}, and \bibinfo{author}{R.~Nandkishore}, \bibinfo{title}{{Localization from Hilbert space shattering: From theory to physical realizations}}, \bibinfo{journal}{\href{http://dx.doi.org/10.1103/PhysRevB.101.174204}{Phys. Rev. B}} \href{http://dx.doi.org/10.1103/PhysRevB.101.174204}{{\bf \bibinfo{volume}{101}}, \bibinfo{pages}{174204}}  (\href{http://dx.doi.org/10.1103/PhysRevB.101.174204}{\bibinfo{year}{2020}}).

\bibitem{PhysRevB.108.045127}
\bibinfo{author}{G.~Francica} and \bibinfo{author}{L.~Dell'Anna}, \bibinfo{title}{{Hilbert space fragmentation in a long-range system}}, \bibinfo{journal}{\href{http://dx.doi.org/10.1103/PhysRevB.108.045127}{Phys. Rev. B}} \href{http://dx.doi.org/10.1103/PhysRevB.108.045127}{{\bf \bibinfo{volume}{108}}, \bibinfo{pages}{045127}}  (\href{http://dx.doi.org/10.1103/PhysRevB.108.045127}{\bibinfo{year}{2023}}).

\bibitem{Moudgalya_2021}
\bibinfo{author}{S.~Moudgalya}, \bibinfo{author}{A.~Prem}, \bibinfo{author}{R.~Nandkishore}, \bibinfo{author}{N.~Regnault}, and \bibinfo{author}{B.~A. Bernevig}, {\em \bibinfo{title}{{Thermalization and Its Absence within Krylov Subspaces of a Constrained Hamiltonian}}\/}, {\em \bibinfo{booktitle}{Memorial Volume for Shoucheng Zhang}\/}, \bibinfo{pages}{147–209} (\bibinfo{publisher}{World Scientific}, \bibinfo{year}{2021}).

\bibitem{PhysRevB.110.045418}
\bibinfo{author}{S.~Aditya}, \bibinfo{author}{D.~Dhar}, and \bibinfo{author}{D.~Sen}, \bibinfo{title}{Subspace-restricted thermalization in a correlated-hopping model with strong hilbert space fragmentation characterized by irreducible strings}, \bibinfo{journal}{\href{http://dx.doi.org/10.1103/PhysRevB.110.045418}{Phys. Rev. B}} \href{http://dx.doi.org/10.1103/PhysRevB.110.045418}{{\bf \bibinfo{volume}{110}}, \bibinfo{pages}{045418}}  (\href{http://dx.doi.org/10.1103/PhysRevB.110.045418}{\bibinfo{year}{2024}}).

\bibitem{PhysRevB.111.045411}
\bibinfo{author}{M.~Ganguli}, \bibinfo{author}{S.~Aditya}, and \bibinfo{author}{D.~Sen}, \bibinfo{title}{Aspects of hilbert space fragmentation in the quantum east model: Fragmentation, subspace-restricted quantum scars, and effects of density-density interactions}, \bibinfo{journal}{\href{http://dx.doi.org/10.1103/PhysRevB.111.045411}{Phys. Rev. B}} \href{http://dx.doi.org/10.1103/PhysRevB.111.045411}{{\bf \bibinfo{volume}{111}}, \bibinfo{pages}{045411}}  (\href{http://dx.doi.org/10.1103/PhysRevB.111.045411}{\bibinfo{year}{2025}}).

\bibitem{PhysRevLett.132.220405}
\bibinfo{author}{P.~\L{}yd\ifmmode~\dot{z}\else \.{z}\fi{}ba}, \bibinfo{author}{P.~Prelov\ifmmode~\check{s}\else \v{s}\fi{}ek}, and \bibinfo{author}{M.~Mierzejewski}, \bibinfo{title}{{Local Integrals of Motion in Dipole-Conserving Models with Hilbert Space Fragmentation}}, \bibinfo{journal}{\href{http://dx.doi.org/10.1103/PhysRevLett.132.220405}{Phys. Rev. Lett.}} \href{http://dx.doi.org/10.1103/PhysRevLett.132.220405}{{\bf \bibinfo{volume}{132}}, \bibinfo{pages}{220405}}  (\href{http://dx.doi.org/10.1103/PhysRevLett.132.220405}{\bibinfo{year}{2024}}).

\bibitem{PhysRevE.104.044106}
\bibinfo{author}{B.~Pozsgay}, \bibinfo{author}{T.~Gombor}, \bibinfo{author}{A.~Hutsalyuk}, \bibinfo{author}{Y.~Jiang}, \bibinfo{author}{L.~Pristy\'ak}, and \bibinfo{author}{E.~Vernier}, \bibinfo{title}{{Integrable spin chain with Hilbert space fragmentation and solvable real-time dynamics}}, \bibinfo{journal}{\href{http://dx.doi.org/10.1103/PhysRevE.104.044106}{Phys. Rev. E}} \href{http://dx.doi.org/10.1103/PhysRevE.104.044106}{{\bf \bibinfo{volume}{104}}, \bibinfo{pages}{044106}}  (\href{http://dx.doi.org/10.1103/PhysRevE.104.044106}{\bibinfo{year}{2021}}).

\bibitem{PhysRevB.101.125126}
\bibinfo{author}{T.~Rakovszky}, \bibinfo{author}{P.~Sala}, \bibinfo{author}{R.~Verresen}, \bibinfo{author}{M.~Knap}, and \bibinfo{author}{F.~Pollmann}, \bibinfo{title}{{Statistical localization: From strong fragmentation to strong edge modes}}, \bibinfo{journal}{\href{http://dx.doi.org/10.1103/PhysRevB.101.125126}{Phys. Rev. B}} \href{http://dx.doi.org/10.1103/PhysRevB.101.125126}{{\bf \bibinfo{volume}{101}}, \bibinfo{pages}{125126}}  (\href{http://dx.doi.org/10.1103/PhysRevB.101.125126}{\bibinfo{year}{2020}}).

\bibitem{PhysRevX.12.011050}
\bibinfo{author}{S.~Moudgalya} and \bibinfo{author}{O.~I. Motrunich}, \bibinfo{title}{{Hilbert Space Fragmentation and Commutant Algebras}}, \bibinfo{journal}{\href{http://dx.doi.org/10.1103/PhysRevX.12.011050}{Phys. Rev. X}} \href{http://dx.doi.org/10.1103/PhysRevX.12.011050}{{\bf \bibinfo{volume}{12}}, \bibinfo{pages}{011050}}  (\href{http://dx.doi.org/10.1103/PhysRevX.12.011050}{\bibinfo{year}{2022}}).

\bibitem{PhysRevB.55.6491}
\bibinfo{author}{S.~Zhang}, \bibinfo{author}{M.~Karbach}, \bibinfo{author}{G.~M\"uller}, and \bibinfo{author}{J.~Stolze}, \bibinfo{title}{{Charge and spin dynamics in the one-dimensional {t-${\mathrm{J}}_{\mathrm{z}}$ and t-J models}}}, \bibinfo{journal}{\href{http://dx.doi.org/10.1103/PhysRevB.55.6491}{Phys. Rev. B}} \href{http://dx.doi.org/10.1103/PhysRevB.55.6491}{{\bf \bibinfo{volume}{55}}, \bibinfo{pages}{6491}}  (\href{http://dx.doi.org/10.1103/PhysRevB.55.6491}{\bibinfo{year}{1997}}).

\bibitem{PhysRevB.36.381}
\bibinfo{author}{C.~Gros}, \bibinfo{author}{R.~Joynt}, and \bibinfo{author}{T.~M. Rice}, \bibinfo{title}{{Antiferromagnetic correlations in almost-localized {Fermi} liquids}}, \bibinfo{journal}{\href{http://dx.doi.org/10.1103/PhysRevB.36.381}{Phys. Rev. B}} \href{http://dx.doi.org/10.1103/PhysRevB.36.381}{{\bf \bibinfo{volume}{36}}, \bibinfo{pages}{381}}  (\href{http://dx.doi.org/10.1103/PhysRevB.36.381}{\bibinfo{year}{1987}}).

\bibitem{Kotrla199033}
\bibinfo{author}{M.~Kotrla}, \bibinfo{title}{{Energy spectrum of the {Hubbard} model with {$U=\infty$}}}, \bibinfo{journal}{\href{http://dx.doi.org/https://doi.org/10.1016/0375-9601(90)90272-P}{Phys. Lett. A}} \href{http://dx.doi.org/https://doi.org/10.1016/0375-9601(90)90272-P}{{\bf \bibinfo{volume}{145}}, \bibinfo{pages}{33}}  (\href{http://dx.doi.org/https://doi.org/10.1016/0375-9601(90)90272-P}{\bibinfo{year}{1990}}).

\bibitem{faddeev1996}
\bibinfo{author}{L.~Faddeev}, {\em \bibinfo{title}{How Algebraic Bethe Ansatz works for integrable model}\/}, {\em \bibinfo{booktitle}{{Fifty Years of Mathematical Physics}}\/}, \bibinfo{pages}{370--439} (\bibinfo{publisher}{World Scientific}, \bibinfo{year}{2016}).

\bibitem{Lanczos_1950}
\bibinfo{author}{C.~Lanczos}, \bibinfo{title}{{{An iteration method for the solution of the eigenvalue problem of linear differential and integral operators}}}, \bibinfo{journal}{\href{http://dx.doi.org/10.6028/jres.045.026}{J. Res. Natl. Bur. Stand. B}} \href{http://dx.doi.org/10.6028/jres.045.026}{{\bf \bibinfo{volume}{45}}, \bibinfo{pages}{255}}  (\href{http://dx.doi.org/10.6028/jres.045.026}{\bibinfo{year}{1950}}).

\bibitem{Park_1986}
\bibinfo{author}{T.~J. Park} and \bibinfo{author}{J.~C. Light}, \bibinfo{title}{{Unitary quantum time evolution by iterative {Lanczos} reduction}}, \bibinfo{journal}{\href{http://dx.doi.org/10.1063/1.451548}{J. Chem. Phys.}} \href{http://dx.doi.org/10.1063/1.451548}{{\bf \bibinfo{volume}{85}}, \bibinfo{pages}{5870}}  (\href{http://dx.doi.org/10.1063/1.451548}{\bibinfo{year}{1986}}).

\bibitem{PhysRevLett.124.040603}
\bibinfo{author}{M.~Mierzejewski} and \bibinfo{author}{L.~Vidmar}, \bibinfo{title}{{Quantitative Impact of Integrals of Motion on the Eigenstate Thermalization Hypothesis}}, \bibinfo{journal}{\href{http://dx.doi.org/10.1103/PhysRevLett.124.040603}{Phys. Rev. Lett.}} \href{http://dx.doi.org/10.1103/PhysRevLett.124.040603}{{\bf \bibinfo{volume}{124}}, \bibinfo{pages}{040603}}  (\href{http://dx.doi.org/10.1103/PhysRevLett.124.040603}{\bibinfo{year}{2020}}).

\bibitem{Mierzejewski_2022}
\bibinfo{author}{M.~Mierzejewski}, \bibinfo{author}{J.~Pawłowski}, \bibinfo{author}{P.~Prelovsek}, and \bibinfo{author}{J.~Herbrych}, \bibinfo{title}{{Multiple relaxation times in perturbed XXZ chain}}, \bibinfo{journal}{\href{http://dx.doi.org/10.21468/scipostphys.13.2.013}{SciPost Phys.}} \href{http://dx.doi.org/10.21468/scipostphys.13.2.013}{{\bf \bibinfo{volume}{13}}, \bibinfo{pages}{013}}  (\href{http://dx.doi.org/10.21468/scipostphys.13.2.013}{\bibinfo{year}{2022}}).

\bibitem{PhysRevB.92.195121}
\bibinfo{author}{M.~Mierzejewski}, \bibinfo{author}{T.~Prosen}, and \bibinfo{author}{P.~Prelov\ifmmode~\check{s}\else \v{s}\fi{}ek}, \bibinfo{title}{{Approximate conservation laws in perturbed integrable lattice models}}, \bibinfo{journal}{\href{http://dx.doi.org/10.1103/PhysRevB.92.195121}{Phys. Rev. B}} \href{http://dx.doi.org/10.1103/PhysRevB.92.195121}{{\bf \bibinfo{volume}{92}}, \bibinfo{pages}{195121}}  (\href{http://dx.doi.org/10.1103/PhysRevB.92.195121}{\bibinfo{year}{2015}}).

\bibitem{PhysRevResearch.5.043019}
\bibinfo{author}{F.~M. Surace} and \bibinfo{author}{O.~Motrunich}, \bibinfo{title}{{Weak integrability breaking perturbations of integrable models}}, \bibinfo{journal}{\href{http://dx.doi.org/10.1103/PhysRevResearch.5.043019}{Phys. Rev. Res.}} \href{http://dx.doi.org/10.1103/PhysRevResearch.5.043019}{{\bf \bibinfo{volume}{5}}, \bibinfo{pages}{043019}}  (\href{http://dx.doi.org/10.1103/PhysRevResearch.5.043019}{\bibinfo{year}{2023}}).

\bibitem{Santos_2010}
\bibinfo{author}{L.~F. Santos} and \bibinfo{author}{M.~Rigol}, \bibinfo{title}{{Onset of quantum chaos in one-dimensional bosonic and fermionic systems and its relation to thermalization}}, \bibinfo{journal}{\href{http://dx.doi.org/10.1103/physreve.81.036206}{Phys. Rev. E}} \href{http://dx.doi.org/10.1103/physreve.81.036206}{{\bf \bibinfo{volume}{81}}}  (\href{http://dx.doi.org/10.1103/physreve.81.036206}{\bibinfo{year}{2010}}).

\bibitem{Fogarty_2021}
\bibinfo{author}{T.~Fogarty}, \bibinfo{author}{M.~A. Garc\'ia-March}, \bibinfo{author}{L.~F. Santos}, and \bibinfo{author}{N.~L. Harshman}, \bibinfo{title}{{Probing the edge between integrability and quantum chaos in interacting few-atom systems}}, \bibinfo{journal}{\href{http://dx.doi.org/10.22331/q-2021-06-29-486}{Quantum}} \href{http://dx.doi.org/10.22331/q-2021-06-29-486}{{\bf \bibinfo{volume}{5}}, \bibinfo{pages}{486}}  (\href{http://dx.doi.org/10.22331/q-2021-06-29-486}{\bibinfo{year}{2021}}).

\bibitem{Szasz_2021}
\bibinfo{author}{D.~Szász-Schagrin}, \bibinfo{author}{B.~Pozsgay}, and \bibinfo{author}{G.~Takacs}, \bibinfo{title}{{Weak integrability breaking and level spacing distribution}}, \bibinfo{journal}{\href{http://dx.doi.org/10.21468/scipostphys.11.2.037}{SciPost Phys.}} \href{http://dx.doi.org/10.21468/scipostphys.11.2.037}{{\bf \bibinfo{volume}{11}}, \bibinfo{pages}{037}}  (\href{http://dx.doi.org/10.21468/scipostphys.11.2.037}{\bibinfo{year}{2021}}).

\bibitem{Sierant_2019}
\bibinfo{author}{P.~Sierant} and \bibinfo{author}{J.~Zakrzewski}, \bibinfo{title}{{Level statistics across the many-body localization transition}}, \bibinfo{journal}{\href{http://dx.doi.org/10.1103/physrevb.99.104205}{Phys. Rev. B}} \href{http://dx.doi.org/10.1103/physrevb.99.104205}{{\bf \bibinfo{volume}{99}}}  (\href{http://dx.doi.org/10.1103/physrevb.99.104205}{\bibinfo{year}{2019}}).

\bibitem{PhysRevB.88.115126}
\bibinfo{author}{Y.~Huang}, \bibinfo{author}{C.~Karrasch}, and \bibinfo{author}{J.~E. Moore}, \bibinfo{title}{{Scaling of electrical and thermal conductivities in an almost integrable chain}}, \bibinfo{journal}{\href{http://dx.doi.org/10.1103/PhysRevB.88.115126}{Phys. Rev. B}} \href{http://dx.doi.org/10.1103/PhysRevB.88.115126}{{\bf \bibinfo{volume}{88}}, \bibinfo{pages}{115126}}  (\href{http://dx.doi.org/10.1103/PhysRevB.88.115126}{\bibinfo{year}{2013}}).

\bibitem{PhysRevE.89.012125}
\bibinfo{author}{M.~Storms} and \bibinfo{author}{R.~R.~P. Singh}, \bibinfo{title}{{Entanglement in ground and excited states of gapped free-fermion systems and their relationship with {Fermi} surface and thermodynamic equilibrium properties}}, \bibinfo{journal}{\href{http://dx.doi.org/10.1103/PhysRevE.89.012125}{Phys. Rev. E}} \href{http://dx.doi.org/10.1103/PhysRevE.89.012125}{{\bf \bibinfo{volume}{89}}, \bibinfo{pages}{012125}}  (\href{http://dx.doi.org/10.1103/PhysRevE.89.012125}{\bibinfo{year}{2014}}).

\bibitem{Beugeling_2015}
\bibinfo{author}{W.~Beugeling}, \bibinfo{author}{A.~Andreanov}, and \bibinfo{author}{M.~Haque}, \bibinfo{title}{{Global characteristics of all eigenstates of local many-body {Hamiltonians}: participation ratio and entanglement entropy}}, \bibinfo{journal}{\href{http://dx.doi.org/10.1088/1742-5468/2015/02/p02002}{J. Stat. Mech.: Theory Exp.}} \href{http://dx.doi.org/10.1088/1742-5468/2015/02/p02002}{{\bf \bibinfo{volume}{2015}}, \bibinfo{pages}{P02002}}  (\href{http://dx.doi.org/10.1088/1742-5468/2015/02/p02002}{\bibinfo{year}{2015}}).

\bibitem{PhysRevE.100.062134}
\bibinfo{author}{T.~LeBlond}, \bibinfo{author}{K.~Mallayya}, \bibinfo{author}{L.~Vidmar}, and \bibinfo{author}{M.~Rigol}, \bibinfo{title}{{Entanglement and matrix elements of observables in interacting integrable systems}}, \bibinfo{journal}{\href{http://dx.doi.org/10.1103/PhysRevE.100.062134}{Phys. Rev. E}} \href{http://dx.doi.org/10.1103/PhysRevE.100.062134}{{\bf \bibinfo{volume}{100}}, \bibinfo{pages}{062134}}  (\href{http://dx.doi.org/10.1103/PhysRevE.100.062134}{\bibinfo{year}{2019}}).

\bibitem{PhysRevLett.119.110604}
\bibinfo{author}{P.~T. Dumitrescu}, \bibinfo{author}{R.~Vasseur}, and \bibinfo{author}{A.~C. Potter}, \bibinfo{title}{{Scaling Theory of Entanglement at the Many-Body Localization Transition}}, \bibinfo{journal}{\href{http://dx.doi.org/10.1103/PhysRevLett.119.110604}{Phys. Rev. Lett.}} \href{http://dx.doi.org/10.1103/PhysRevLett.119.110604}{{\bf \bibinfo{volume}{119}}, \bibinfo{pages}{110604}}  (\href{http://dx.doi.org/10.1103/PhysRevLett.119.110604}{\bibinfo{year}{2017}}).

\bibitem{PhysRevResearch.6.023030}
\bibinfo{author}{J.~\ifmmode~\check{S}\else \v{S}\fi{}untajs}, \bibinfo{author}{M.~Hopjan}, \bibinfo{author}{W.~De~Roeck}, and \bibinfo{author}{L.~Vidmar}, \bibinfo{title}{{Similarity between a many-body quantum avalanche model and the ultrametric random matrix model}}, \bibinfo{journal}{\href{http://dx.doi.org/10.1103/PhysRevResearch.6.023030}{Phys. Rev. Res.}} \href{http://dx.doi.org/10.1103/PhysRevResearch.6.023030}{{\bf \bibinfo{volume}{6}}, \bibinfo{pages}{023030}}  (\href{http://dx.doi.org/10.1103/PhysRevResearch.6.023030}{\bibinfo{year}{2024}}).

\bibitem{You_2007}
\bibinfo{author}{W.-L. You}, \bibinfo{author}{Y.-W. Li}, and \bibinfo{author}{S.-J. Gu}, \bibinfo{title}{{Fidelity, dynamic structure factor, and susceptibility in critical phenomena}}, \bibinfo{journal}{\href{http://dx.doi.org/10.1103/physreve.76.022101}{Phys. Rev. E}} \href{http://dx.doi.org/10.1103/physreve.76.022101}{{\bf \bibinfo{volume}{76}}}  (\href{http://dx.doi.org/10.1103/physreve.76.022101}{\bibinfo{year}{2007}}).

\bibitem{Kolodrubetz_2017}
\bibinfo{author}{M.~Kolodrubetz}, \bibinfo{author}{D.~Sels}, \bibinfo{author}{P.~Mehta}, and \bibinfo{author}{A.~Polkovnikov}, \bibinfo{title}{{Geometry and non-adiabatic response in quantum and classical systems}}, \bibinfo{journal}{\href{http://dx.doi.org/10.1016/j.physrep.2017.07.001}{Phys. Rep.}} \href{http://dx.doi.org/10.1016/j.physrep.2017.07.001}{{\bf \bibinfo{volume}{697}}, \bibinfo{pages}{1}}  (\href{http://dx.doi.org/10.1016/j.physrep.2017.07.001}{\bibinfo{year}{2017}}).

\bibitem{Kim_2024}
\bibinfo{author}{H.~Kim} and \bibinfo{author}{A.~Polkovnikov}, \bibinfo{title}{{Integrability as an attractor of adiabatic flows}}, \bibinfo{journal}{\href{http://dx.doi.org/10.1103/physrevb.109.195162}{Phys. Rev. B}} \href{http://dx.doi.org/10.1103/physrevb.109.195162}{{\bf \bibinfo{volume}{109}}}  (\href{http://dx.doi.org/10.1103/physrevb.109.195162}{\bibinfo{year}{2024}}).

\bibitem{D_Alessio_2016}
\bibinfo{author}{L.~D’Alessio}, \bibinfo{author}{Y.~Kafri}, \bibinfo{author}{A.~Polkovnikov}, and \bibinfo{author}{M.~Rigol}, \bibinfo{title}{{From quantum chaos and eigenstate thermalization to statistical mechanics and thermodynamics}}, \bibinfo{journal}{\href{http://dx.doi.org/10.1080/00018732.2016.1198134}{Adv. Phys.}} \href{http://dx.doi.org/10.1080/00018732.2016.1198134}{{\bf \bibinfo{volume}{65}}, \bibinfo{pages}{239}}  (\href{http://dx.doi.org/10.1080/00018732.2016.1198134}{\bibinfo{year}{2016}}).

\bibitem{Brenes_2020b}
\bibinfo{author}{M.~Brenes}, \bibinfo{author}{T.~LeBlond}, \bibinfo{author}{J.~Goold}, and \bibinfo{author}{M.~Rigol}, \bibinfo{title}{{Eigenstate Thermalization in a Locally Perturbed Integrable System}}, \bibinfo{journal}{\href{http://dx.doi.org/10.1103/physrevlett.125.070605}{Phys. Rev. Lett.}} \href{http://dx.doi.org/10.1103/physrevlett.125.070605}{{\bf \bibinfo{volume}{125}}}  (\href{http://dx.doi.org/10.1103/physrevlett.125.070605}{\bibinfo{year}{2020}}).

\bibitem{Brenes_2020}
\bibinfo{author}{M.~Brenes}, \bibinfo{author}{J.~Goold}, and \bibinfo{author}{M.~Rigol}, \bibinfo{title}{{Low-frequency behavior of off-diagonal matrix elements in the integrable {XXZ }chain and in a locally perturbed quantum-chaotic {XXZ} chain}}, \bibinfo{journal}{\href{http://dx.doi.org/10.1103/physrevb.102.075127}{Phys. Rev. B}} \href{http://dx.doi.org/10.1103/physrevb.102.075127}{{\bf \bibinfo{volume}{102}}}  (\href{http://dx.doi.org/10.1103/physrevb.102.075127}{\bibinfo{year}{2020}}).

\bibitem{PhysRevE.102.062113}
\bibinfo{author}{T.~LeBlond} and \bibinfo{author}{M.~Rigol}, \bibinfo{title}{{Eigenstate thermalization for observables that break {Hamiltonian} symmetries and its counterpart in interacting integrable systems}}, \bibinfo{journal}{\href{http://dx.doi.org/10.1103/PhysRevE.102.062113}{Phys. Rev. E}} \href{http://dx.doi.org/10.1103/PhysRevE.102.062113}{{\bf \bibinfo{volume}{102}}, \bibinfo{pages}{062113}}  (\href{http://dx.doi.org/10.1103/PhysRevE.102.062113}{\bibinfo{year}{2020}}).

\bibitem{Vidmar_2021}
\bibinfo{author}{L.~Vidmar}, \bibinfo{author}{B.~Krajewski}, \bibinfo{author}{J.~Bonča}, and \bibinfo{author}{M.~Mierzejewski}, \bibinfo{title}{{Phenomenology of Spectral Functions in Disordered Spin Chains at Infinite Temperature}}, \bibinfo{journal}{\href{http://dx.doi.org/10.1103/physrevlett.127.230603}{Phys. Rev. Lett.}} \href{http://dx.doi.org/10.1103/physrevlett.127.230603}{{\bf \bibinfo{volume}{127}}}  (\href{http://dx.doi.org/10.1103/physrevlett.127.230603}{\bibinfo{year}{2021}}).

\bibitem{lydzba2024}
\bibinfo{author}{P.~\L{}yd\ifmmode~\dot{z}\else \.{z}\fi{}ba}, \bibinfo{author}{R.~\ifmmode \acute{S}\else \'{S}\fi{}wi\ifmmode~\mbox{\k{e}}\else \k{e}\fi{}tek}, \bibinfo{author}{M.~Mierzejewski}, \bibinfo{author}{M.~Rigol}, and \bibinfo{author}{L.~Vidmar}, \bibinfo{title}{Normal weak eigenstate thermalization}, \bibinfo{journal}{\href{http://dx.doi.org/10.1103/PhysRevB.110.104202}{Phys. Rev. B}} \href{http://dx.doi.org/10.1103/PhysRevB.110.104202}{{\bf \bibinfo{volume}{110}}, \bibinfo{pages}{104202}}  (\href{http://dx.doi.org/10.1103/PhysRevB.110.104202}{\bibinfo{year}{2024}}).

\bibitem{lisiecki_2025_17434301}
\bibinfo{author}{M.~Lisiecki}, \bibinfo{author}{J.~Bonca}, \bibinfo{author}{M.~Mierzejewski}, \bibinfo{author}{J.~Herbrych}, and \bibinfo{author}{P.~Łydżba}, \bibinfo{title}{Dataset for "{T}unable {H}ilbert space fragmentation and extended critical regime"}  (\bibinfo{year}{2025}). \bibinfo{note}{Zenodo, \url{https://zenodo.org/records/17434301}}.

\bibitem{Srednicki_1999}
\bibinfo{author}{M.~Srednicki}, \bibinfo{title}{{The approach to thermal equilibrium in quantized chaotic systems}}, \bibinfo{journal}{\href{http://dx.doi.org/10.1088/0305-4470/32/7/007}{J. Phys. A Math. Gen.}} \href{http://dx.doi.org/10.1088/0305-4470/32/7/007}{{\bf \bibinfo{volume}{32}}, \bibinfo{pages}{1163}}  (\href{http://dx.doi.org/10.1088/0305-4470/32/7/007}{\bibinfo{year}{1999}}).

\bibitem{Sch_nle_2021}
\bibinfo{author}{C.~Schönle}, \bibinfo{author}{D.~Jansen}, \bibinfo{author}{F.~Heidrich-Meisner}, and \bibinfo{author}{L.~Vidmar}, \bibinfo{title}{{Eigenstate thermalization hypothesis through the lens of autocorrelation functions}}, \bibinfo{journal}{\href{http://dx.doi.org/10.1103/physrevb.103.235137}{Phys. Rev. B}} \href{http://dx.doi.org/10.1103/physrevb.103.235137}{{\bf \bibinfo{volume}{103}}}  (\href{http://dx.doi.org/10.1103/physrevb.103.235137}{\bibinfo{year}{2021}}).

\bibitem{PhysRevE.87.012118}
\bibinfo{author}{R.~Steinigeweg}, \bibinfo{author}{J.~Herbrych}, and \bibinfo{author}{P.~Prelov\ifmmode~\check{s}\else \v{s}\fi{}ek}, \bibinfo{title}{{Eigenstate thermalization within isolated spin-chain systems}}, \bibinfo{journal}{\href{http://dx.doi.org/10.1103/PhysRevE.87.012118}{Phys. Rev. E}} \href{http://dx.doi.org/10.1103/PhysRevE.87.012118}{{\bf \bibinfo{volume}{87}}, \bibinfo{pages}{012118}}  (\href{http://dx.doi.org/10.1103/PhysRevE.87.012118}{\bibinfo{year}{2013}}).

\bibitem{PhysRevE.91.012144}
\bibinfo{author}{W.~Beugeling}, \bibinfo{author}{R.~Moessner}, and \bibinfo{author}{M.~Haque}, \bibinfo{title}{{Off-diagonal matrix elements of local operators in many-body quantum systems}}, \bibinfo{journal}{\href{http://dx.doi.org/10.1103/PhysRevE.91.012144}{Phys. Rev. E}} \href{http://dx.doi.org/10.1103/PhysRevE.91.012144}{{\bf \bibinfo{volume}{91}}, \bibinfo{pages}{012144}}  (\href{http://dx.doi.org/10.1103/PhysRevE.91.012144}{\bibinfo{year}{2015}}).

\end{thebibliography}

\end{document}